\documentclass[preprint,nofootinbib,aps,superscriptaddress,eqsecnum]{revtex4-2} 
\usepackage{float}
\pdfoutput=1
\usepackage{extarrows}
\usepackage{tabularx}
\usepackage{graphicx}
\usepackage{amsmath}
\usepackage{amsfonts}
\usepackage{amssymb}
\usepackage{hyperref}
\usepackage[justification=centering]{caption} 
\usepackage[margin=1.5cm]{geometry}
\usepackage{graphicx,subcaption,lipsum}
\let\revappendix\appendix
\usepackage{caption}
\usepackage{subcaption}
\usepackage{color}
\allowdisplaybreaks
\def\bea{\begin{eqnarray}}
	\def\eea{\end{eqnarray}}
\def\be{\begin{equation}}
	\def\ee{\end{equation}}

\begin{document}
	\title{Leptogenesis, $0\nu\beta\beta$ and lepton flavor violation in modular left-right asymmetric model with polyharmonic $Maa\beta$ forms}
	\author{Bhabana Kumar}
	\email{bhabana12@tezu.ernet.in}
	\author{Mrinal Kumar Das}
	\email{mkdas@tezu.ernet.in}
	\affiliation{Department of Physics, Tezpur University, Tezpur 784028, India}

	\begin{abstract}
		In the absence of supersymmetry, modular forms need not be holomorphic functions of the modulus $\tau$. Using this idea, we construct a non-supersymmetric framework using polyharmonic $Maa\beta$ forms. In this approach, the Yukawa coupling is no longer strictly holomorphic in $\tau$ but instead incorporates both holomorphic and non-holomorphic components. We realize a non-supersymmetric, left-right asymmetric model based on the $\Gamma_3$ modular group, where the active neutrino masses are generated via an extended inverse seesaw mechanism. The model successfully predicts the sum of neutrino masses below the current experimental bound and accommodates neutrino mixing angles within the $3\sigma$ range. Given its strong predictive power in neutrino oscillation parameters, we further explore its implications for beyond Standard Model (BSM) phenomena, including neutrinoless double beta ($0\nu\beta\beta$) decay, lepton flavor violation (LFV), and baryogenesis via leptogenesis (BAU). Our findings indicate that the model predicts an effective Majorana mass and LFV branching ratios consistent with experimental constraints while also providing a viable explanation for the observed baryon asymmetry through resonant leptogenesis.
	\end{abstract}
	
	\maketitle
	\newpage
\section{\textbf{Introduction}}
The neutrino is an enigmatic particle. The Standard Model (SM) predicts that neutrinos are massless due to the absence of their right-handed counterparts. However, the discovery of neutrino oscillation\cite{Bellini:2013wra} confirmed that neutrinos have small but nonzero and non-degenerate masses. This discovery suggested the need to extend the SM by introducing new particles. The seesaw mechanism\cite{Minkowski:1977sc,Mohapatra:1979ia,Rodejohann:2004qh}, the left-right symmetric model\cite{Pati:1974yy,Senjanovic:1975rk,Deshpande:1990ip,Mohapatra:1974gc}, and Grand Unified Theories are some proposed extensions aimed at addressing the limitations of the SM.\\
Another significant problem that needs to be addressed is the baryon asymmetry of the universe. In 1967, Andrei Sakharov proposed three necessary conditions, known as the Sakharov conditions\cite{Sakharov:1967dj}, for the dynamical generation of baryon asymmetry. These conditions are:  
	\begin{enumerate}
		\item Baryon number violation,  
		\item C and CP violation, and  
		\item Out-of-equilibrium.  
	\end{enumerate}  
	
This asymmetry is quantified by a parameter, $\eta$, which represents the ratio of baryon to photon density. Cosmological observations estimate this ratio to be on the order of $10^{-10}$\cite{ParticleDataGroup:2018ovx}.  
\begin{equation}\label{W2O1}
		\eta= \frac{n_{B}-n_{\bar{B}}}{n_{\gamma}} = (6.12\pm 0.04)\times 10^{-10}
\end{equation}
SM satisfies all three Sakharov conditions. However, considering the CP violation present within the SM, it is found that the quantity $\eta$ is of the order of $10^{-20}$, which is very small compared to the observed value, indicating the need for a new source of CP violation. Many mechanisms have been proposed to explain the observed baryon asymmetry of the universe, and one such mechanism is leptogenesis, where lepton asymmetry is first produced by the out-of-equilibrium decay of heavy neutrinos. This lepton asymmetry is then converted into baryon asymmetry through processes called sphalerons\cite{Trodden:2004mj,Rubakov:1996vz}. There are different types of leptogenesis, such as thermal leptogenesis and resonant leptogenesis. In this work, we have used the resonant leptogenesis process to study the baryon asymmetry of the universe.
There is another phenomenon people are also interested in, which is the $0\nu\beta\beta$ decay, and its existence will provide insight into the ongoing issue of whether the neutrino is a Dirac or Majorana particle. In $0\nu\beta\beta$ decay, two protons within the nucleus change into two neutrons by emitting two electrons. The process can be represented as follows:  
\begin{equation*}
		X^{A}_{(Z,N)} \rightarrow X^{A}_{(Z-2,N+2)} + 2e^{-}
\end{equation*}
The decay is only possible if the neutrino is its own antiparticle, implying the Majorana nature of the neutrino. Different experiments like GERDA\cite{GERDA:2020xhi}, KamLAND-Zen\cite{KamLAND-Zen:2016pfg}, and CUORE have been attempting to detect signals for $0\nu\beta\beta$ decay. Current experiments have provided bounds on the effective mass and half-life of $0\nu\beta\beta$ decay. According to these experiments, the effective mass should be less than $0.165-0.065$ eV, and the half-life should be greater than $10^{26}$ years. The bound on the effective mass appears as a range due to uncertainties in calculating nuclear matrix elements.\\
The existence of massive neutrinos suggests the possibility of charged  LFV. Within the SM, it is possible to give neutrinos a nonzero mass by extending the SM with three right-handed (RH) neutrinos, where the neutrino mass is generated via the Higgs mechanism. With this consideration, we can have two-body and three-body decays of charged leptons. However, the branching ratio for such decays is extremely small, of the order of $10^{-50}$. Due to this small branching ratio, these decays within the SM are not accessible to current experiments. However, other neutrino mass generation mechanisms, such as the seesaw mechanism and models like GUTs, predict an enhanced branching ratio that falls within the experimental sensitivity range. In this work, we have calculated the branching ratios for the decay processes $\mu \rightarrow e^{-}+\gamma$, $\tau \rightarrow \mu + \gamma$, and $\tau \rightarrow e^{-}+\gamma$. The bounds on the branching ratios for these three processes are given in Table \ref{W2T4}.\\
To study neutrino masses and mixing, we have constructed a model using the $\Gamma_{3}$ modular group. The advantage of using modular symmetry over discrete flavour symmetry is that the model can be constructed without any flavon field. Models built using discrete flavour symmetry require an extra field known as a "flavon" to break the flavour symmetry, and the predictivity of the model significantly depends on the vacuum alignment of the flavon field. However, in the case of modular symmetry, the flavour symmetry is broken using the modulus $\tau$\cite{Feruglio:2017spp,Novichkov:2019sqv,King:2020qaj}, thereby minimizing the need for flavon fields. Modular symmetry has been used in models based on supersymmetry, where the Yukawa couplings are holomorphic functions of the modulus $\tau$. However, in a non-supersymmetric framework, it is possible for the Yukawa couplings to have non-holomorphic components as well. The present work focuses on constructing a non-supersymmetric neutrino mass model, where the Yukawa couplings are polyharmonic functions of $\tau$, known as polyharmonic $Maa\beta$ forms\cite{Qu:2024rns,Ding:2020zxw}. Another advantage of using the polyharmonic $Maa\beta$ form is that we can incorporate zero and negative modular weights in this framework. This provides a broader area for exploration compared to the simple modular symmetry approach, where only positive and nonzero modular weights can be used as modular forms. A detailed discussion regarding the modular group and polyharmonic $Maa\beta$ form is given in Appendix \ref{W2A1}.  The study is mainly focused on constructing a non-supersymmetric model using the $\Gamma_{3}$ modular group and validating the model by comparing its predictions with current neutrino oscillation data. Additionally, we examine its capability in explaining other phenomena such as $0\nu\beta\beta$ decay, LFV, and the BAU. We have calculated the sum of neutrino masses and mixing angles from the model. To analyze the model's contribution to $0\nu\beta\beta$ decay and LFV, we have computed the effective mass due to the $\lambda$ contribution and evaluated the branching ratios for different LFV processes. Finally, we have calculated the baryon asymmetry of the universe using resonant leptogenesis.\\
The overall paper is structured as follows: In Section \ref{W2S1}, we provide a brief discussion on the left-right asymmetric model and the extended inverse seesaw mechanism. Section \ref{W2S2} describes the model we have constructed. Sections \ref{W2S3},\ref{W2S4}, and \ref{W2S5} focus on LFV, $0\nu\beta\beta$ decay, and resonant leptogenesis, respectively. In these sections, we also present the relevant formulas used in the calculations of the branching ratio, effective mass, and CP asymmetry term for the LFV process, $0\nu\beta\beta$ decay, and BAU, respectively. Finally, Sections \ref{W2S6} and \ref{W2S7} provide a detailed discussion on the numerical analysis, results, and conclusions of our study.
\section{\textbf{Left-Right Asymmetric model and Extended inverse seesaw mechanism}}\label{W2S1}
The limitation of the SM is overcome by adding more particles or by introducing a new gauge group. One such model is the Left-Right symmetric model (LRSM). The gauge group for LRSM is $SU(2)_{L}\times SU(2)_{R}\times U(1)_{B-L}\times D$. where D is the discrete parity symmetry. In LRSM both the D-parity and the $SU(2)_{R}$ gauge group break on the same energy scale and we get the same gauge coupling for the gauge group $SU(2)_{L}$ and $SU(2)_{R}$. But there is another possibility to decouple the D parity from the $SU(2)_{R}$ group which is by giving a non-zero vacuum expectation value to a scalar field\cite{Chang:1983fu,Rizzo:1981dm}. In such case, D-parity breaks at a high energy scale but the $SU(2)_{R}$ gauge group breaks at some lower energy scale. Due to the spontaneous symmetry breaking of the D-parity at a high energy scale, we observed unequal gauge coupling values for the gauge group $SU(2)_{L}$ and $SU(2)_{R}$. This alternative way of breaking D-parity is known as the left-right asymmetric model\cite{Chang:1984uy}. The $SU(2)_{R}$ gauge group can again break down to $U(1)_{R}$ by giving non-zero V.E.V to a scalar triplet. The entire symmetry-breaking steps are given below.
	\begin{center} 
		\centering
		$ SU(2)_{L} \times SU(2)_{R} \times U(1)_{B-L} \times  SU(3)_{C} \times D $ \\ $ \downarrow $ \hspace{0.5cm} $\sigma$ \\ 
		
		$ SU(2)_{L} \times SU(2)_{R} \times U(1)_{B-L} \times SU(3)_{C}$ \\      
		$\downarrow $ \vspace{0.5cm} $\Sigma$\\ 
		
		$ SU(2)_{L} \times U(1)_{R} \times U(1)_{B-L} \times SU(3)_{C}$ \\ $ \downarrow$ \vspace{0.5 cm} $\Delta_{R}$\\ 
		
		$ SU(2)_{L} \times U(1)_{Y} \times SU(3)_{C}  $ \\
		$\downarrow $\hspace{0.5cm} $\Phi$ \\
		
		$ U(1)_{em} \times SU(3)_{C}$. \\  
		
	\end{center}
In the present work, we consider the gauge group $SU(2)_{L} \times U(1)_{R} \times U(1)_{B-L}$ to construct the model and generate the light neutrino mass matrix using the extended inverse seesaw mechanism. In LRSM, the RH neutrinos and charged right-handed leptons transform as a doublet under the $ SU(2)_{R}$ gauge group. However, in this model, we assume that the $ SU(2)_{R}$ gauge group breaks down to $U(1)_{R}$. Consequently, the right-handed charged leptons and neutrinos no longer transform as a doublet but instead, become singlets and carry a $U(1)_{R}$ charge, which corresponds to the third component of the isospin of $SU(2)_{R}$. To implement the extended inverse seesaw mechanism, we introduce one sterile fermion per generation and one scalar doublet into the model. The particle content of the model and its charge assignments under the gauge group are given in Table \ref{W2T2}.  
The Lagrangian for the model is given in the equation \eqref{W2Q2}
	\begin{equation}\label{W2Q2}
		\mathcal{L} = Y^{l}\bar{\Psi}_{L}N_{R}\Phi + Y^{f}N^{C}_{R}N_{R}\Delta_{R} + Y^{t}\bar{N}_{R}S\chi_{R} + S^{T}\mu_{S}S + h.c~ .
	\end{equation}
After the electroweak symmetry breaking of this Lagrangian gives rise to a $9\times 9$ matrix in the basis $(\nu_{L}, N_{R}, S)^{T}$.
	\begin{equation}\label{W2Q4}
	\mathcal{M}=	\begin{pmatrix}
			0 & M_{D} & 0 \\
			M^{T}_{D} & M_{R} & M_{N}^{T} \\
			0 & M_{N} & M_{S}
		\end{pmatrix} ~~.
	\end{equation}
where $M_{D}$ and $M_{R}$ are the Dirac and Majorana mass matrices, which are given by $M_{D} = Y^{l} \langle \Phi \rangle$ and $M_{R} = Y^{f} \langle \Delta_{R} \rangle$, respectively. Similarly, $M_{S}$ represents the sterile-sterile mass matrix, and $M_{N} = Y^{t} \langle \chi_{R} \rangle$ denotes the RH neutrino and sterile ($N-S$) mixing matrix. The block-diagonalization of this $9 \times 9$ mass matrix results in the mass matrices for the light neutrinos, sterile neutrinos, and RH neutrinos, as given in equation \eqref{W2Q5}.
	\begin{gather} \label{W2Q5}
		\begin{aligned}
			m_{\nu} & = M_{D}M_{N}^{-1}M_{S}(M_{D}M_{N}^{-1})^{T}  \\ 
			m_{S}  &= M_{S} - M_{N}M^{-1}_{R}M_{N}^{T}  \\ 
			m_{R} &= M_{R}~ .
		\end{aligned}
	\end{gather}
To obtain the eigenvalues we further diagonalized the  matrices given in equation \eqref{W2Q5} by their respective unitary matrices as follows
	\begin{gather} \label{W2Q6}
		\begin{aligned}
			\hat{m}_{\nu} & = U^{\dagger}_{\nu}m_{\nu}U_{\nu}^{*} = diag(m_{\nu_{1}}, m_{\nu_{2}}, m_{\nu_{3}})\\
			\hat{m}_{S} & =U^{\dagger}_{S}m_{s}U^{*}_{S} = diag(m_{S_{1}}, m_{S_{2}}, m_{S_{3}}) \\
			\hat{m}_{R} & = U^{\dagger}_{N}m_{R}U^{*}_{N} = diag(m_{R_{1}}, m_{R_{2}}, m_{R_{3}})~~~. \\
		\end{aligned}
	\end{gather}
The complete mixing matrix responsible for diagonalizing the $9\times 9$ mass matrix given in the equation  \ref{W2Q4} is shown bellow \cite{Senapati:2020alx,Awasthi:2013ff,Grimus:2000vj}
	\begin{align}\label{W2Q7}
		\mathbf{V}=\begin{pmatrix}
			V^{\nu\nu}& V^{\nu S} & V^{\nu N} \\
			V^{S\nu} & V^{SS} & V^{SN} \\
			V^{N\nu} & V^{NS} & V^{NN} 
		\end{pmatrix}=
		\begin{pmatrix}
			(1-\frac{1}{2}XX^{\dagger})U_{\nu} & (X-\frac{1}{2}ZY^{\dagger})U_{S}& ZU_{N} \\
			-X^{\dagger}U_{\nu} & (1-\frac{1}{2}(X^{\dagger}X+YY^{\dagger}))U_{S} & (Y-\frac{1}{2}X^{\dagger}Z)U_{N}\\
			y^{*}X^{\dagger}U_{\nu} & -Y^{\dagger}U_{S} & (1-\frac{1}{2}Y^{\dagger}Y)U_{N}
		\end{pmatrix} 
	\end{align}
	where $X=M_{D}M_{N}^{-1},~ Y=M_{N}M_{R}^{-1},~Z=M_{D}M^{-1}_{R},~$ and $y=M_{N}^{-1}M_{S}$ \\
We know that the active neutrino mass matrix can be diagonalized by a unitary matrix known as the PMNS matrix ($U_{PMNS}$). The matrix $U_{PMNS}$ can be parametrized by using three mixing angles and three phases, among which one is a Dirac CP phase denoted as $\delta_{CP}$ and two are Majorana phases ($\alpha$, $\beta$). By using standard parametrization, we can write the $U_{PMNS}$ matrix in the following way
	\begin{align}\label{W2Q8}
		U_{PMNS}= \begin{pmatrix}
			c_{12}c_{13} & s_{12}c_{13} & s_{13}e^{-i\delta_{CP}} \\
			-s_{12}c_{23}-c_{12}s_{23}s_{13}e^{i\delta_{CP}} & c_{12}c_{23}-s_{12}s_{23}s_{13}e^{i\delta_{CP}} & s_{23}c_{13} \\
			s_{12}s_{23}-c_{12}c_{23}s_{13} e^{i\delta_{CP}} & -c_{12}s_{23}-s_{12}c_{23}s_{13}e^{i\delta_{CP}} & c_{23}c_{13} \\
		\end{pmatrix}.
		\begin{pmatrix}
			1 & 0 & 0\\
			0 & e^{i\alpha} & 0\\
			0 & 0 & e^{i\beta}
		\end{pmatrix}~.
	\end{align}
Where $c_{ij}$ and $s_{ij}$ stands for $\cos\theta_{ij}$ and $\sin\theta_{ij}$, respectively. One can represent all the mixing angles in terms of the elements of the $U_{PMNS}$ matrix as given in equation \eqref{W2Q9}, and their $3\sigma$ values of neutrino oscillation parameters are given in Table \ref{W2T1}. We have used those $ 3\sigma$ ranges given in Table \ref{W2T1} for our numerical analysis part.
\begin{table}[H]
		\centering
		\begin{tabular}{|c|c|c|} \hline
			\textbf{Oscillation Parameter} & \textbf{Normal Hierarchy} & \textbf{Inverted Hierarchy} \\ \hline
			
			$\sin^{2}\theta_{12}$ & $0.275 \rightarrow 0.345$ & $0.275 \rightarrow 0.345$ \\ [4ex] \hline
			$\sin^{2}\theta_{23}$ & $0.435 \rightarrow 0.585$ & $0.440 \rightarrow 0.585$ \\ [4ex] \hline
			$\sin^{2}\theta_{13}$ & $0.02030 \rightarrow 0.02388$ & $0.02060 \rightarrow 0.02409$ \\ [4ex] \hline
			$\delta_{CP}$ & $124 \rightarrow 364$ & $201 \rightarrow 335$ \\ [4ex] \hline
			$\frac{\Delta m_{21}^{2}}{10^{-5}}$ & $6.92 \rightarrow 8.05$ & $6.92 \rightarrow 8.05$ \\ [4ex] \hline
			$\frac{\Delta m_{3l}^{2}}{10^{-3}}$ & $+2.451 \rightarrow 2.578$ & $-2.547 \rightarrow -2.421$\\ [4ex] \hline
		\end{tabular}
		\caption{$3\sigma$ values of oscillation parameters~\cite{Esteban:2024eli}}
		\label{W2T1}
\end{table}
	\begin{align}\label{W2Q9}
		\sin^{2}\theta_{13} = |(U_{PMNS})_{13}|^{2} ~, ~~ \sin^{2}\theta_{23} = \frac{|(U_{PMNS})_{23}|^{2}}{1-|(U_{PMNS})_{13}|^{2}} ~, ~~ \sin^{2}\theta_{12}=\frac{|(U_{PMNS})_{12}|^{2}}{1-|(U_{PMNS})_{13}|^{2}}
	\end{align}
\section{\textbf{The Model}}\label{W2S2}
The light neutrino mass is generated via an extended inverse seesaw mechanism. To implement this mechanism, an extra singlet sterile particle $S_{i}$ per generation and one scalar doublet $\chi_{R}$ are added to the model\cite{Senapati:2020alx,Awasthi:2013ff,Parida:2012sq}. The charges assigned under the $A_{4}$ group and the modular weights of the particle content of the model are given in Table \ref{W2T2}.~ Table \ref{W2T3} provides information regarding the modular weights and charges under $A_{4}$ for the Yukawa couplings. The charge assignments are determined such that each term in the Lagrangian is a trivial singlet under the $A_{4}$ group.\\
	\begin{table}[ht]
		\centering
		\begin{tabular}{|m{1.5cm}|c| c|c|c|c|c|c|}
			\hline
			
			Field &$\Psi_{R_{i}}$ &$ \Psi_{L_{i}} $ & $N_{R} $ & $  S$ & $\Phi$ & $\Delta_{R}$ & $\chi_{R}$  \\ \hline
			
			$ SU(2)_{L}$& 1& 2 & 1 & 1 & 2 & 1 & 1 \\ \hline
			
			$U(1)_R$ &$-\frac{1}{2}$ &0 & $\frac{1}{2}$ & 0 &$-\frac{1}{2}$ & -1 &$\frac{1}{2}$ \\ \hline
			
			$ A_{4} $ & 1,$1^{\prime\prime}$,$1^{\prime}$  &1,$1^{\prime}$,$1^{\prime\prime}$ & 3 & 3 & 1 & 1 & 1 \\ \hline
			
			$k_{I}$ & 0 & $ 0 $ & 0 & 0 & 0 & 0 & 0 \\ \hline
		\end{tabular}
		\caption{Charge assignment for the particle content of the model}
		\label{W2T2}
	\end{table}
	\begin{table}[ht]
		\centering
		\begin{tabular}{|c|c|c|} \hline
			& Polyharmonic $Maa\beta$ ($Y^{k}_{r}$)\\ \hline
			$A_{4}$ & $3, 1$ \\ \hline
			$k_{I}$ & $0$ \\ \hline 
		\end{tabular}
		\caption {Charge assignment and modular weight for Yukawa coupling}
		\label{W2T3}
	\end{table}
The Lagrangian associated with the model is given in equation \ref{W2Q10}
\begin{equation}\label{W2Q10}
		\mathcal{L} =\mathcal{L}_{l} + \mathcal{L}_{D} +\mathcal{L}_{M} +\mathcal{L}_{N-S} +\mathcal{L}_{S} + h.c ~~.
	\end{equation}
The first two terms in the Lagrangian represent the Yukawa interaction terms for the charged leptons, while the second and third terms represent the Dirac and Majorana terms for the neutral leptons, respectively. The fourth term describes the mixing between the RH and sterile neutrinos ($N-S$), and finally, the last term represents the coupling term for the sterile neutrino. By using the charge assignments and modular weights under the $A_{4}$ group, we have constructed the invariant Yukawa Lagrangian for the model. From the Lagrangian, we have derived the mass matrices in terms of modular forms. To obtain a diagonal charged lepton mass matrix, we have considered the left-handed lepton doublets ($\Psi_{L_{i}}$) to transform as $(1, 1^{\prime}, 1^{\prime\prime})$ under the $A_{4}$ group, and the right-handed charged leptons ($\Psi_{R_{i}}$) to transform as $(1, 1^{\prime\prime}, 1^{\prime})$ under the $A_{4}$ group. The final Yukawa interaction term between the charged leptons and the diagonal charged lepton mass matrix is given in equations~\eqref{W2Q11} and~\eqref{W2Q12}, respectively.
\begin{equation}\label{W2Q11}  
		\mathcal{L}_{l} = \alpha_{1} \Phi \bar{\Psi}_{L_{1}} Y^{0}_{1}\Psi_{R_{1}} + \alpha_{2} \Phi \bar{\Psi}_{L_{2}} Y^{0}_{1}\Psi_{R_{2}}+\alpha_{3} \Phi \bar{\Psi}_{L_{3}} Y^{0}_{1}\Psi_{R_{3}}  
	\end{equation} 
	\begin{equation}\label{W2Q12}  
		M_{l}= v \begin{pmatrix}  
			Y^{0}_{1}\alpha_{1} & 0 & 0 \\  
			0 & Y^{0}_{1}\alpha_{2} & 0 \\  
			0 & 0 & Y^{0}_{1}\alpha_{3}  
		\end{pmatrix} ~~ .  
		\end{equation}
Where $Y^{0}_{1}$ and all the other parameters in the mass matrix are constants so we can adjust their values to obtain the required charged lepton mass matrix.\\
The $A_{4}$ invariant Dirac term is presented in equation \eqref{W2Q13} and the corresponding Dirac mass matrix is given in equation \eqref{W2Q14}
\begin{equation}\label{W2Q13}
		\mathcal{L}_{D} =q_{1}\Phi \bar{\Psi}_{L_{1}}(Y^{(0)}_{3}N_{R})_{1}+q_{2}\Phi \bar{\Psi}_{L_{2}}(Y^{(0)}_{3}N_{R})_{1^{\prime\prime}}+q_{3}\Phi \bar{\Psi}_{L_{3}}(Y^{(0)}_{3}N_{R})_{1^{\prime}}~~.
\end{equation}
\begin{equation}\label{W2Q14}
	M_{D}= v \begin{pmatrix}
		q_{1} Y^{(0)}_{3,1} & q_{1}Y^{(0)}_{3,3} & q_{1} Y^{(0)}_{3,2} \\
		q_{2}Y^{(0)}_{3,3} & q_{2}Y^{(0)}_{3,2} & q_{2}Y^{(0)}_{3,1} \\
		q_{3}Y^{(0)}_{3,2} & q_{3}) Y^{(0)}_{3,1} & q_{3}Y^{(0)}_{3,3}
	\end{pmatrix} ~~ .
\end{equation}
Where $q_{1}$, $q_{2}$ and $q_{3}$ are adjustable parameters of the model. $v$ is the VEV of the bidoublet $\Phi$.
The $A_{4}$ invariant Majorana mass term can be constructed in the following way 
\begin{equation}\label{W2Q15}
		\mathcal{L}_{M_{R}} = q_{4}\Delta_{R} Y^{(0)}_{3}(\bar{N^{C}}_{R}N_{R})_{3_S} +  q_{5}\Delta_{R} Y^{(0)}_{1}(\bar{N^{C}}_{R}N_{R})_{1} ~~ .
\end{equation}
From the above Yukawa lagrangian term we have constructed the Majorana mass mass matrix, which is given in equation~\eqref{W2Q16}
\begin{equation}\label{W2Q16}
	\begin{aligned}
		M_{R} &=  v_{R}\begin{pmatrix}
			2 q_{4} Y^{(0)}_{3,1} & -q_{4}Y^{(0)}_{3,3} & - q_{4}Y^{(0)}_{3,2} \\
			-q_{4}Y^{(0)}_{3,3} & 2q_{4} Y^{(0)}_{3,2} & -q_{4}Y^{(0)}_{3,1} \\
			-q_{4}Y^{(0)}_{3,2} & -q_{4}Y^{(0)}_{3,1} & 2 q_{4}Y^{(0)}_{3,3}
		\end{pmatrix}+ v_{R} \begin{pmatrix}
		q_{5} & 0 & 0 \\
		0 & 0 & q_{5} \\
		0 & q_{5} & 0
		\end{pmatrix}
		 \\
		&= v_{R} \begin{pmatrix}
			q_{5}+2 q_{4} Y^{(0)}_{3,1} & -q_{4}Y^{(0)}_{3,3} & - q_{4}Y^{(0)}_{3,2} \\
			-q_{4}Y^{(0)}_{3,3} & 2q_{4} Y^{(0)}_{3,2} & q_{5}-q_{4}Y^{(0)}_{3,1} \\
			-q_{4}Y^{(0)}_{3,2} & q_{5}-g_{4}Y^{(0)}_{3,1} & 2 q_{4}Y^{(0)}_{3,3}
		\end{pmatrix}
	\end{aligned}
\end{equation}
Similarly, we have constructed the $A_{4}$ invariant $N-S$ and sterile coupling term, which is given in the equation~ \eqref{W2Q17} and \eqref{W2Q19} respectively and their corresponding mixing matrix are given in equation~\eqref{W2Q18} and \eqref{W2Q20}\\

\begin{equation}\label{W2Q17}
	\mathcal{L}_{N} = q_{6} \chi_{R} (\bar{N}_{R} S)_{3_{S}} Y^{(0)}_{3} 
	+ q_{7} \chi_{R} (\bar{N}_{R} S)_{3_{A}} Y^{(0)}_{3} 
	+ q_{8} \chi_{R} (\bar{N}_{R} S)_{1} Y^{(0)}_{1} \, .
\end{equation}
\begin{equation}\label{W2Q18}
	\begin{aligned}
		M_{N} &=  v^{\prime}q_{6}\begin{pmatrix}
			2 Y^{(0)}_{3,1} & -Y^{(0)}_{3,3} & -Y^{(0)}_{3,2} \\
			-Y^{(0)}_{3,3} & 2Y^{(0)}_{3,2} & -Y^{(0)}_{3,1} \\
			-Y^{(0)}_{3,2} & -Y^{(0)}_{3,1} & 2 Y^{(0)}_{3,3}
		\end{pmatrix} +  v^{\prime}q_{7}\begin{pmatrix}
		0 & Y^{(0)}_{3,3} & - Y^{(0)}_{3,2} \\
		-Y^{(0)}_{3,3} & 0 & Y^{(0)}_{3,1} \\
		Y^{(0)}_{3,2} & -Y^{(0)}_{3,1} & 0
		\end{pmatrix}+ v^{\prime}q_{8} \begin{pmatrix}
			1 & 0 & 0 \\
			0 & 0 & 1 \\
			0 & 1 & 0
		\end{pmatrix}
		\\
		&=  v^{\prime} \begin{pmatrix}
			q_{8} + 2q_{6} Y^{(0)}_{3,1} & (-q_{6} + q_{7}) Y^{(0)}_{3,3} & (-q_{6} - q_{7}) Y^{(0)}_{3,2} \\
			(-q_{6} - q_{7}) Y^{(0)}_{3,3} & 2q_{6} Y^{(0)}_{3,2} & q_{8} + (-q_{6} + q_{7}) Y^{(0)}_{3,1} \\
			(-q_{6} + q_{7}) Y^{(0)}_{3,2} & q_{8} + (-q_{6} - q_{7}) Y^{(0)}_{3,1} & 2q_{6} Y^{(0)}_{3,3}
		\end{pmatrix} \, .
	\end{aligned}
\end{equation}

\begin{equation}\label{W2Q19}
	\mathcal{L}_S = g_{1} Y^{(0)}_{1}(S^{T}S) + g_{2} Y^{(0)}_{3} (S^{T} S)_{3_{S}} \, .
\end{equation}

\begin{equation}\label{W2Q20}
	\begin{aligned}
		M_{S} &=  g_{1}\begin{pmatrix}
			1 & 0 & 0 \\
			0 & 0 & 1 \\
			0 & 1 & 0
		\end{pmatrix}
		+ g_{2} \begin{pmatrix}
			2 Y^{(0)}_{3,1} & - Y^{(0)}_{3,3} & - Y^{(0)}_{3,2} \\
			- Y^{(0)}_{3,3} & 2 Y^{(0)}_{3,2} & - Y^{(0)}_{3,1} \\
			- Y^{(0)}_{3,2} & - Y^{(0)}_{3,1} & 2 Y^{(0)}_{3,3}
		\end{pmatrix} \\[10pt]
		&= \begin{pmatrix}
			g_{1} + 2g_{2} Y^{(0)}_{3,1} & -g_{2} Y^{(0)}_{3,3} & -g_{2} Y^{(0)}_{3,2} \\
			-g_{2} Y^{(0)}_{3,3} & 2g_{2} Y^{(0)}_{3,2} & g_{1} - g_{2} Y^{(0)}_{3,1} \\
			-g_{2} Y^{(0)}_{3,2} & g_{1} - g_{2} Y^{(0)}_{3,1} & 2g_{2} Y^{(0)}_{3,3}
		\end{pmatrix}
	\end{aligned}
\end{equation}
$q_{i}$ ($i=1$ to $8$) and $g_{j}$ ($j=1,2$) are the adjustable parameters of the model, and they are complex numbers. $\Phi=246$ GeV is the VEV of the bidoublet. $v_{R}$ and $v^{\prime}$ are the VEVs of the scalar triplet $\Delta_{R}$ and the scalar doublet $\chi_{R}$, respectively, and their values lie in the TeV range. Finally, $q_{9}$ and $q_{10}$ are taken in the keV range. \\
After constructing the Dirac, Majorana, $N-S$, and sterile mass matrices in terms of modular forms, we use equation \eqref{W2Q5} to calculate the mass matrices of the light neutrino, sterile neutrino, and RH neutrino in terms of the Yukawa couplings.
\section{\textbf{Neutrinoless double beta decay}}\label{W2S4}
In the left-right asymmetric model, along with the standard contribution, we also have non-standard contributions to the $0\nu\beta\beta$ decay. The contributions come from the $W_{L}-W_{L}$, $W_{R}-W_{R}$, $W_{L}-W_{R}$ currents and also from the charged Higgs bosons. In our previous work, we have already discussed the effective mass and half-life of the $0\nu\beta\beta$,  decay for the $W_{L}-W_{L}$ and $W_{R}-W_{R}$ currents. In this work, we calculated the effective mass due to the $W_{L}-W_{R}$ current. In the case of $W_{L}-W_{R}$, we can have two types of mixed helicity Feynman diagrams, which is known as the $\lambda$ and $\eta$ diagrams. The formula to calculate the effective mass for those two contributions due to the exchange of light neutrino, heavy RH, and sterile neutrino is given in the equations ~ \eqref{W2Q23} and \eqref{W2Q24} respectively.
	\begin{itemize}
		\item Contribution to $\lambda$ diagram
		
		\begin{gather}\label{W2Q23}
			\begin{aligned}
				m_{ee,\lambda}^{\nu} &= 10^{-2}\Big(\frac{M_{W_{L}}}{M_{W_{R}}}\Big)^{2} \Big(\frac{g_{R}}{g_{L}}\Big)\sum^{3}_{i=1} V_{ei}^{\nu\nu}V_{ei}^{N\nu}|p| , \\ 
				m_{ee,\lambda}^{N} &= 10^{-2}\Big(\frac{M_{W_{L}}}{M_{W_{R}}}\Big)^{2} \Big(\frac{g_{R}}{g_{L}}\Big)\sum^{3}_{i=1} V_{ei}^{\nu N}V_{ei}^{N N} \frac{|p|^{3}}{M^{2}_{S_{i}}}, \\ 
				m_{ee,\lambda}^{S} &= 10^{-2}\Big(\frac{M_{W_{L}}}{M_{W_{R}}}\Big)^{2} \Big(\frac{g_{R}}{g_{L}}\Big)\sum^{3}_{i=1} V_{ei}^{\nu S}V_{ei}^{N S}\frac{|p|^{3}}{M^{2}_{S_{i}}}.
			\end{aligned}
		\end{gather}
		\item Contribution due to $\eta$ diagram
		\begin{gather}\label{W2Q24}
			\begin{aligned}
				m_{ee,\eta}^{\nu} &=\Big(\frac{M_{W_{L}}}{M_{W_{R}}}\Big)^{2} \Big(\frac{g_{R}}{g_{L}}\Big)\sum^{3}_{i=1} V_{ei}^{\nu\nu}V_{ei}^{N\nu}|p| , \\ 
				m_{ee,\eta}^{N} &= \Big(\frac{M_{W_{L}}}{M_{W_{R}}}\Big)^{2} \Big(\frac{g_{R}}{g_{L}}\Big)\sum^{3}_{i=1} V_{ei}^{\nu N}V_{ei}^{N N}\tan\zeta \frac{|p|^{3}}{M^{2}_{S_{k}}}\\
				m_{ee,\eta}^{S} &= \Big(\frac{M_{W_{L}}}{M_{W_{R}}}\Big)^{2} \Big(\frac{g_{R}}{g_{L}}\Big)\sum^{3}_{i=1} V_{ei}^{\nu S}V_{ei}^{N S}\tan \zeta \frac{|p|^{3}}{M^{2}_{S_{k}}}.
			\end{aligned}
		\end{gather}
		
	\end{itemize}
\section{\textbf{Lepton Flavor Violating Processes}}\label{W2S3}
The $9\times 9$ matrix given in the equation \eqref{W2Q4} can be diagonalised by a $9\times 9$ matrix. If $\mathbf{V}$ is the diagonalizing matrix then we can write\\
\begin{equation}\label{W2Q21}
 \mathbf{V}^{\dagger} \mathcal{M}\mathbf{V}= diag(m_{ii},\Psi_{jj}) 
\end{equation}
where $i=1$ to $3$ and $j=4$ to $9$. In the obtained diagonal matrix the first three eigenvalues represent the mass of the light neutrino and the remaining six eigenvalues represent the mass of the heavy eigenstates. The presence of heavy mass eigenstates in the model contributes to the LFV process and the formula for calculating the branching ratio is given below\cite{Deppisch:2004fa}
\begin{equation}\label{W2Q22}
		Br(l_{\alpha}\rightarrow l_{\beta}+\gamma)= \frac{\alpha^{3}_{w}\sin^{2}\theta_{w}m^{5}_{l_{\alpha}}}{256 \pi^{2}M^{4}_{W}\Gamma_{\alpha}}\big|\mathcal{G}^{N}_{\alpha\beta} + \mathcal{G^{S}_{\alpha\beta}}\big|^{2} 
	\end{equation}
where \begin{gather*}
    \begin{aligned}
	\mathcal{G}^{N}_{\alpha\beta} &= \sum_{k}\Big(V^{\nu N}\big)_{\alpha k} \big(V^{\nu N}\big)^{*}_{\beta k} \mathcal{I}\Big(\frac{m^{2}_{N_{k}}}{M^{2}_{W_{L}}}\Big) , \\ 
	\mathcal{G}^{S}_{\alpha\beta} &= \sum_{j}\Big(V^{\nu S}\big)_{\alpha j} \big(V^{\nu S}\big)^{*}_{\beta j} \mathcal{I}\Big(\frac{m^{2}_{S_{j}}}{M^{2}_{W_{L}}}\Big) , \\ 
	\mathcal{I}(x) &= -\frac{2x^{3}+5x^{2}-x}{4(1-x)^{3}}- \frac{3x^{3}\ln{x}}{2(1-x)^{4}} .
\end{aligned}
	\end{gather*}
In equation \eqref{W2Q22}, $\alpha_{w}$, $\theta_{w}$, and $M_{W}$ represent the fine structure constant, Weinberg angle, and mass of the right-handed gauge boson respectively. $m_{l_{\alpha}}$, and  $\Gamma_{\alpha}$ is the mass and decay width of the lepton $l_{\alpha}$ respectively.  $(V^{\nu S})_{\alpha k}$ and $(V^{\nu N})_{\alpha j}$ are the elements of diagonalizing matrix $\mathbf{V}$.  
	\begin{table}[H]
		\centering
		\begin{tabular}{|c|c|c|}
			\hline
			Branching ratio for LFV processes & Experimental bounds \\ \hline
			$ Br(\tau \rightarrow e\gamma)$ & $< 1.5 \times 10^{-8}$ \cite{BaBar:2009hkt}  \\ \hline
			$Br(\tau \rightarrow \mu\gamma)$ & $< 1.5 \times 10^{-8}$ \cite{BaBar:2009hkt} \\ \hline
			$Br(\mu \rightarrow e\gamma)$ & $<4.2 \times 10^{-13}$ \cite{MEG:2016leq} \\ \hline
		\end{tabular}
		\caption{Experimental upper bound on LFV process}
		\label{W2T4}
	\end{table}	
\section{\textbf{Resonent Leptogensis}}\label{W2S5}
For resonant leptogenesis, the primary requirement is the existence of a quasi-Dirac pair. In this mechanism, the lepton asymmetry is generated by the decay of the lightest quasi-Dirac pair, with their mass splitting being comparable to their decay width. Within the model, there exist six heavy mass eigenstates, comprising three RH neutrinos and three sterile neutrinos. The mass matrix for these heavy eigenstates is given by
	\begin{equation}\label{W2Q25}
		M=\begin{pmatrix}
			M_{R} & M_{N}^{T} \\
			M_{N} & M_{S}
		\end{pmatrix},
	\end{equation}
which is a $6\times6$ matrix. This matrix is diagonalized using a unitary matrix $\mathcal{V}$, yielding the mass spectrum for the six heavy neutrinos, three of which correspond to heavy RH neutrinos, while the remaining three are heavy sterile neutrinos. The diagonalized mass matrix is expressed as
	\begin{equation}\label{W2Q26}
		M_{\text{diag}}=\mathcal{V}^{T}M\mathcal{V}=\text{diag}\Big(M_{\Psi_{jj}}\Big),
	\end{equation}
where $j$ runs from $1$ to $6$. The CP asymmetry term is computed on the basis where $M$ is diagonal. On this basis, the Lagrangian in Equation \eqref{W2Q2} can be rewritten as
	\begin{equation}
		\mathcal{L}^{\prime} = h_{i\alpha}\bar{\Psi_{i}}\Phi \Psi_{L} +  M_{\text{diag}}\Psi^{T}_{i}C^{-1}\Psi_{i} + h.c,
	\end{equation}
where $h_{i\alpha}$ represents the Yukawa couplings in the mass basis. The relationship between the Yukawa couplings in the mass basis ($h_{i\alpha}$) and those in the flavor basis ($Y^{l}$) is given by\cite{Chakraborty:2021azg,Blanchet:2010kw,Agashe:2018cuf}
	\begin{gather} \label{W2Q27}
		\begin{aligned}
			h_{1\alpha} & = Y_{1\alpha}^{l}V_{11}^{*}+Y_{2\alpha}^{l}V_{12}^{*}+Y_{3\alpha}^{l}V_{13}^{*}, \\
			h_{2\alpha}  &= Y_{1\alpha}^{l}V_{21}^{*}+Y_{2\alpha}^{l}V_{22}^{*}+Y_{3\alpha}^{l}V_{23}^{*}, \\
			h_{3\alpha} &= Y_{1\alpha}^{l}V_{31}^{*}+Y_{2\alpha}^{l}V_{32}^{*}+Y_{3\alpha}^{l}V_{33}^{*}, \\
			h_{4\alpha} &= Y_{1\alpha}^{l}V_{41}^{*}+Y_{2\alpha}^{l}V_{42}^{*}+Y_{3\alpha}^{l}V_{43}^{*}, \\
			h_{5\alpha} &= Y_{1\alpha}^{l}V_{51}^{*}+Y_{2\alpha}^{l}V_{52}^{*}+Y_{3\alpha}^{l}V_{53}^{*}, \\
			h_{6\alpha} &= Y_{1\alpha}^{l}V_{61}^{*}+Y_{2\alpha}^{l}V_{62}^{*}+Y_{3\alpha}^{l}V_{63}^{*}.
		\end{aligned}
	\end{gather}
The CP asymmetry term $\epsilon_{k}$, corresponding to the decay of heavy neutrinos into leptons and Higgs bosons ($\Psi_{i}\rightarrow \Psi_{L}\Phi$), is expressed as\cite{Covi:1996wh}
	\begin{equation}\label{W2Q28}
		\epsilon_{k}= \frac{\sum_{\alpha}\Big[\Gamma(\Psi_{i}\rightarrow \Psi_{L}\phi)-\Gamma(\Psi_{i}\rightarrow\bar{\Psi_{L}}\Phi^{\dagger})\Big]}{\sum_{\alpha}\Big[\Gamma(\Psi_{i}\rightarrow \Psi_{L}\phi)+\Gamma(\Psi_{i}\rightarrow\bar{\Psi_{L}}\Phi^{\dagger})\Big]} =\frac{1}{8\pi}\sum_{k\neq i}\frac{\text{Im}\Big[(hh^{\dagger})_{ik}^{2}\Big]}{(hh^{\dagger})_{kk}}f_{ik},
	\end{equation}
	where $f_{ik}=\frac{(M^{2}_{i}-M^{2}_{k})M_{i}M_{k}}{(M^{2}_{i}-M^{2}_{k})^{2}+R^{2}_{ik}}$\cite{Garny:2011hg,Iso:2013lba,Iso:2014afa}. Here, $R_{ik}=|M_{i}\Gamma_{i}+M_{k}\Gamma_{k}|$, and $\Gamma_{j}=\frac{(hh^{\dagger})_{jj}M_{j}}{8\pi}$ denotes the total decay width of $\Psi_{j}$.
 The CP asymmetry is given by
	\begin{gather} \label{W2Q29}
		\begin{aligned}
			\epsilon_{1} & =\frac{1}{8\pi (hh^{\dagger})_{11}}\text{Im}\Big[(hh^{\dagger})^{2}_{12}f_{12}+(hh^{\dagger})^{2}_{13}f_{13}+(hh^{\dagger})^{2}_{14}f_{14}+(hh^{\dagger})^{2}_{15}f_{15}+(hh^{\dagger})^{2}_{16}f_{16}\Big] \\
			\epsilon_{2} &= \frac{1}{8\pi (hh^{\dagger})_{22}}\text{Im}\Big[(hh^{\dagger})^{2}_{21}f_{21}+(hh^{\dagger})^{2}_{23}f_{23}+(hh^{\dagger})^{2}_{24}f_{24}+(hh^{\dagger})^{2}_{25}f_{25}+(hh^{\dagger})^{2}_{26}f_{26}\Big]\\
			\epsilon_{3} & =\frac{1}{8\pi (hh^{\dagger})_{33}}\text{Im}\Big[(hh^{\dagger})^{2}_{31}f_{31}+(hh^{\dagger})^{2}_{32}f_{32}+(hh^{\dagger})^{2}_{34}f_{34}+(hh^{\dagger})^{2}_{35}f_{35}+(hh^{\dagger})^{2}_{36}f_{36}\Big] \\
			\epsilon_{4} &= \frac{1}{8\pi (hh^{\dagger})_{44}}\text{Im}\Big[(hh^{\dagger})^{2}_{41}f_{41}+(hh^{\dagger})^{2}_{42}f_{42}+(hh^{\dagger})^{2}_{43}f_{43}+(hh^{\dagger})^{2}_{45}f_{45}+(hh^{\dagger})^{2}_{46}f_{46}\Big]\\
			\epsilon_{5} & =\frac{1}{8\pi (hh^{\dagger})_{55}}\text{Im}\Big[(hh^{\dagger})^{2}_{51}f_{51}+(hh^{\dagger})^{2}_{52}f_{52}+(hh^{\dagger})^{2}_{53}f_{53}+(hh^{\dagger})^{2}_{54}f_{54}+(hh^{\dagger})^{2}_{56}f_{56}\Big] \\
			\epsilon_{6} &= \frac{1}{8\pi (hh^{\dagger})_{66}}\text{Im}\Big[(hh^{\dagger})^{2}_{61}f_{61}+(hh^{\dagger})^{2}_{62}f_{62}+(hh^{\dagger})^{2}_{63}f_{63}+(hh^{\dagger})^{2}_{64}f_{64}+(hh^{\dagger})^{2}_{65}f_{65}\Big].
		\end{aligned}
	\end{gather}
Since the asymmetry produced by the heavy pair is subject to washout effects, the washout parameter in terms of the Hubble parameter is given by
	\begin{equation}\label{W2Q30}
		K_{i}=\frac{\Gamma_{i}}{H}=\frac{M_{i}}{8\pi}(hh^{\dagger})_{ii} \frac{M_{\text{pl}}}{1.66\sqrt{g^{*}}M^{2}_{i}},
	\end{equation}
	where $M_{\text{pl}}=1.2\times10^{19} GeV $ represents the Planck mass and $g_{*}$ denotes the effective number of degrees of freedom. The final expression for BAU is given by
	\begin{equation}\label{W2Q31}
	 \eta=10^{-2}\sum d_{i}\epsilon_{k},
	\end{equation}
	where $d_{i}$ is the dilution factor responsible for the washout of the asymmetry associated with the heavy pair. The value of the dilution factor depends on the range of the washout parameter, and the expressions for the dilution factor corresponding to different ranges of the washout parameter are given below\cite{Branco:2002kt}\\
	\begin{equation} \label{W2Q32}
		-d \approx 
		\begin{cases}
			\displaystyle \sqrt{0.1\,K} \exp\left(-\frac{4}{3(0.1\,K)^{0.25}} \right), & \text{for } K \geq 10^{6}, \\[10pt]
			\displaystyle \frac{0.3}{K \ln(K)^{0.6}}, & \text{for } 10 < K < 10^{6}, \\[10pt]
			\displaystyle \frac{1}{2\sqrt{K^{2}+9}}, & \text{for } 0 \leq K \leq 10.
		\end{cases}
		\end{equation}
\section{\textbf{Numarical Analysis}}\label{W2S6}
The free parameters are considered as complex numbers and are adjusted such that most of the calculated values of the sum of the neutrino masses ($\sum m_{\nu}$) and mixing angles calculated from the model fall within the experimental bounds.  Additionally, these ranges allow the model to consistently explain all the studied phenomena while remaining within current experimental bounds. The range of these parameters is given below:  
	\[
	\text{Re}[q_{i}] = \text{Im}[q_{i}] \in [0.01,0.1]
	\]
Additionally, we have taken \( v_{R} = v^{\prime} = 10 \) TeV and $ g_{1}=g_{2} \in [10, 50] $ keV.
From this point onward, we denote the three Yukawa couplings $Y^{(0)}_{3,1}$, $Y^{(0)}_{3,2}$, and $Y^{(0)}_{3,3}$ as $Y_{1}$, $Y_{2}$, and $Y_{3}$, respectively. This notation is used consistently throughout the paper.
\subsection{Parameter space of Yukawa coupling}
\begin{figure}[H]
		\centering
        \begin{subfigure}[b]{0.45\textwidth}
			\centering
			\includegraphics[width=9cm, height=6cm]{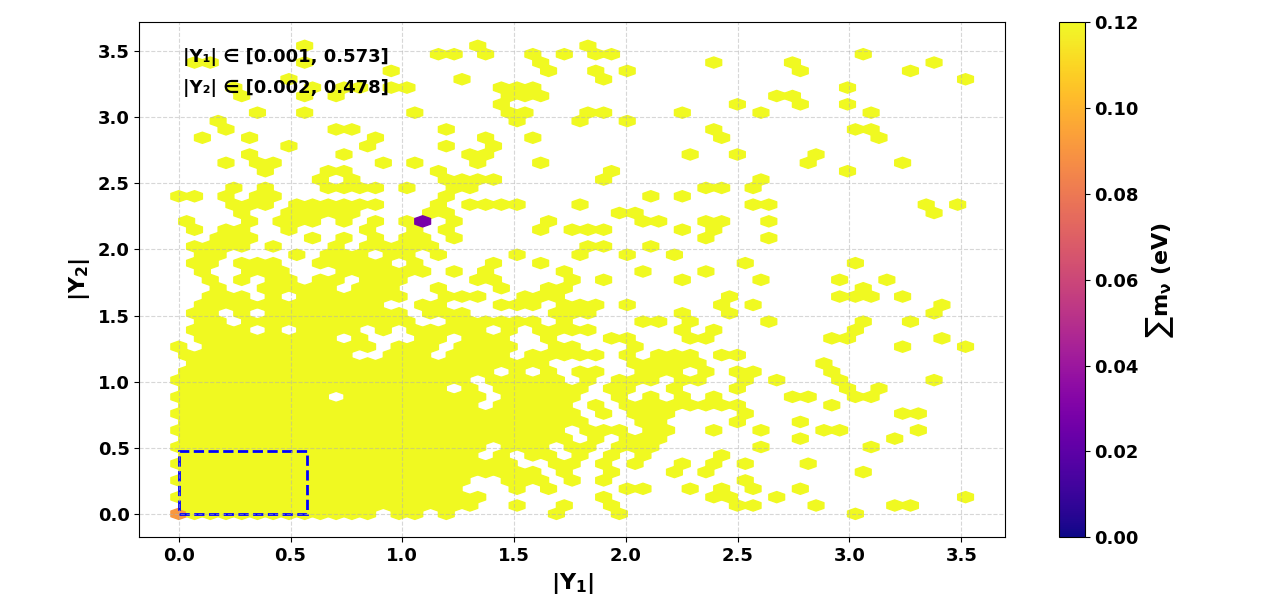}
			\caption{For NH}
			\label{p2f1}
		\end{subfigure}
		\hfill
		\begin{subfigure}[b]{0.45\textwidth}
			\centering
			\includegraphics[width=9cm, height=6cm]{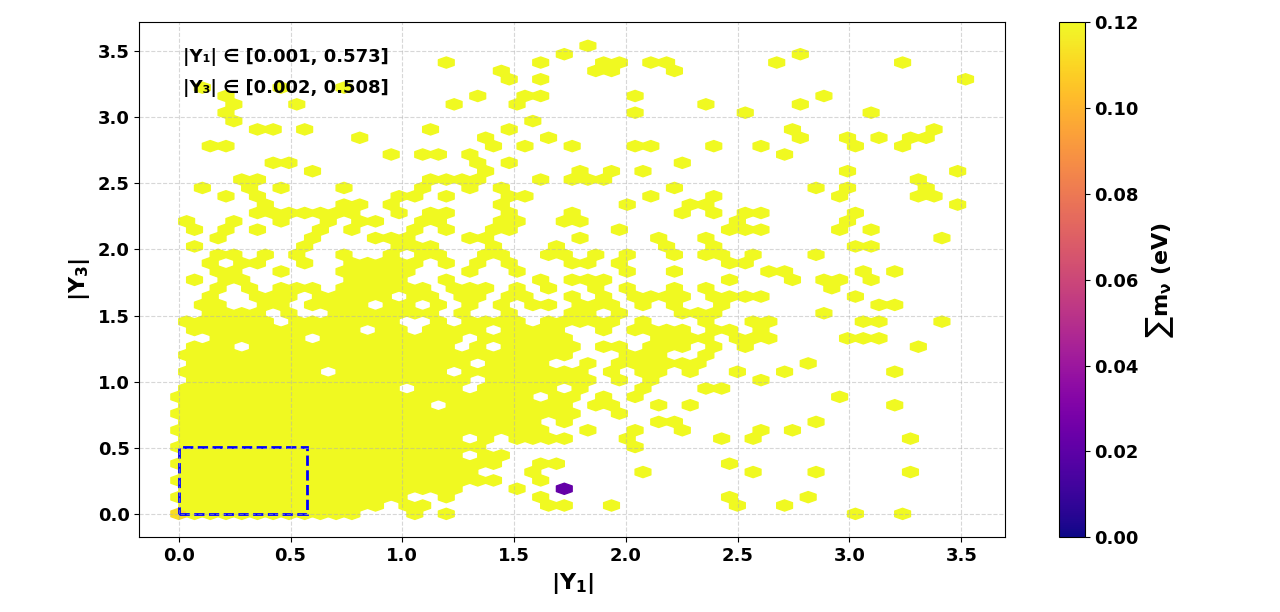}
			\caption{For NH}
			\label{p2f2}
		\end{subfigure}
	\begin{subfigure}[b]{0.45\textwidth}
			\centering
			\includegraphics[width=9cm, height=6cm]{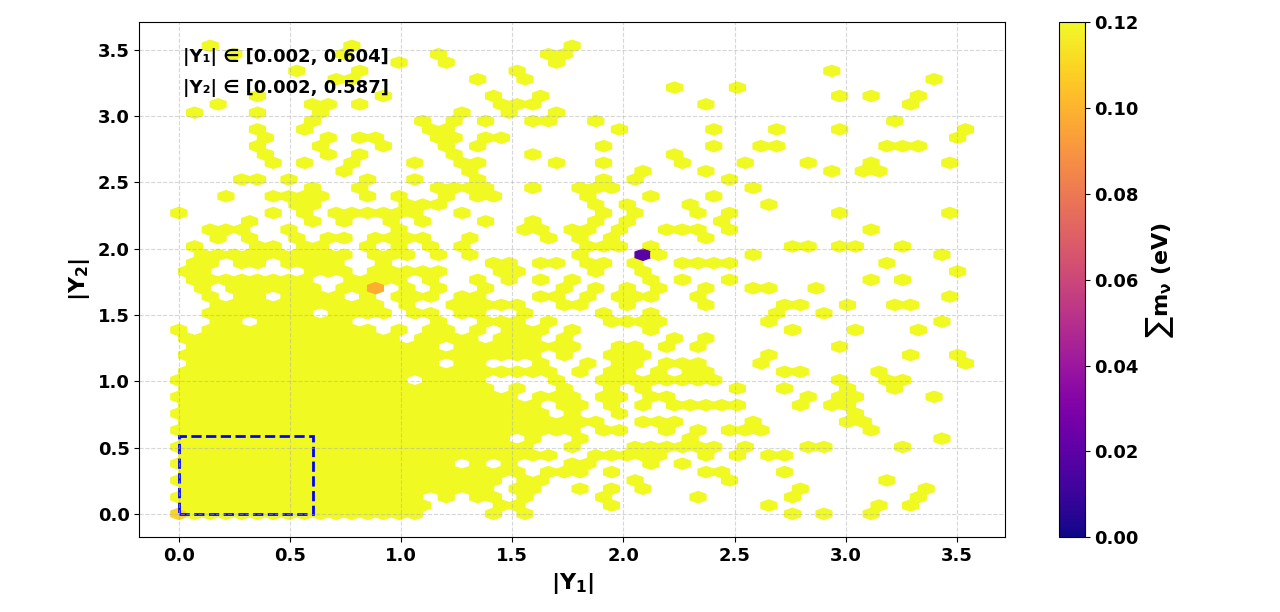}
			\caption{For IH}
			\label{p2f3}
		\end{subfigure}
		\hfill
		\begin{subfigure}[b]{0.45\textwidth}
			\centering
			\includegraphics[width=9cm, height=6cm]{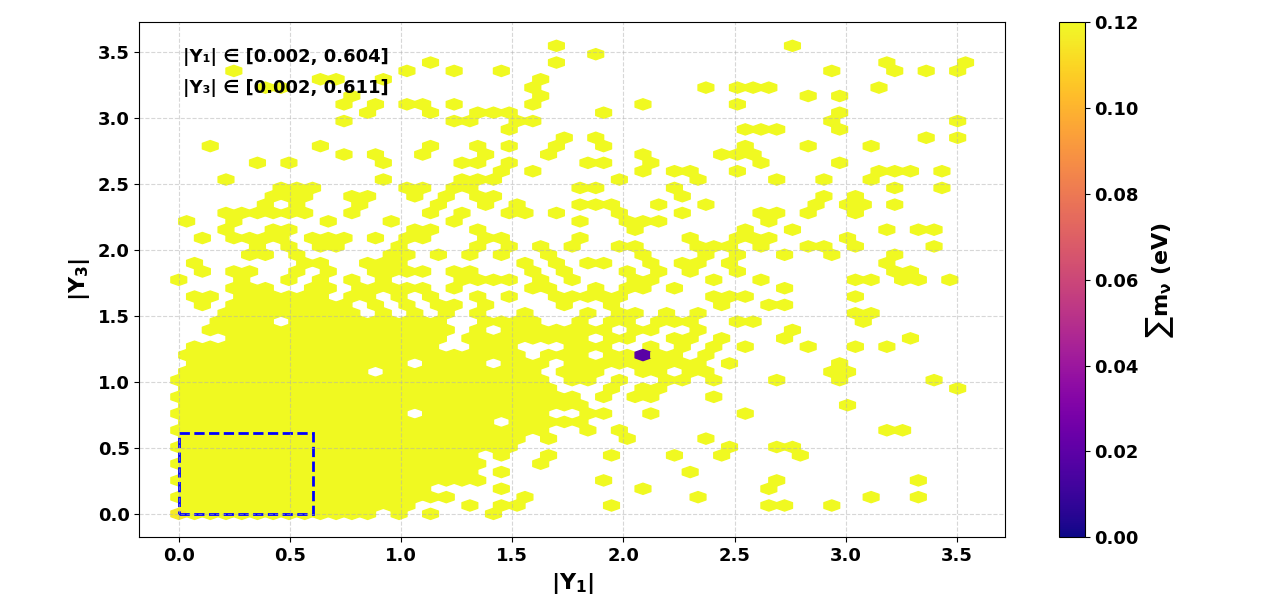}
			\caption{For IH}
			\label{p2f4}
		\end{subfigure}
		\caption{Figure shows the parameter space of Yukawa couplings that satisfy the Planck bound on the $\sum m_{\nu}$ for NH and IH}
		\label{P2F1a}
	\end{figure}

The light neutrino mass matrix has been constructed using the matrices derived from the model, following Equation \eqref{W2Q5}. Utilizing the $3\sigma$ values of the neutrino oscillation parameters which is given in Table \ref{W2T1}, we have calculated the Yukawa couplings associated with the model. After determining the values of the three Yukawa couplings, we computed the $\sum m_{\nu}$ and mixing angles and compared these calculated values with the experimental ranges. To analyze the parameter space, we have plotted density plots to identify the regions of Yukawa couplings where all neutrino oscillation parameters fall within the $3\sigma$ range. Since the model contains three Yukawa couplings ($Y_{1}$, $Y_{2}$, and $Y_{3}$), we have considered the Yukawa couplings as independent variables and the neutrino oscillation parameters as dependent variables in the density plots. Similar density plots have also been generated for the IH. In all the plots, the rectangular box highlights the region containing 90\% of the data points. Within the same plots, we also indicate the preferred range of Yukawa couplings corresponding to this 90\% distribution. Density plots have been generated for the three mixing angles, namely $\theta_{12}$, $\theta_{23}$, and $\theta_{13}$, to analyze their dependence on the Yukawa couplings. From those plots, we have identified a common range for the three Yukawa couplings which keep the mixing angles within the $3\sigma$ range for both NH and IH. Similarly, Figure ~\ref{P2F1a} shows the variation of $\sum m_{\nu} $ with the Yukawa couplings for NH and IH. After determining the parameter space for the Yukawa couplings that satisfy the neutrino oscillation constraints, we have used only these values in the calculations of the effective Majorana mass, branching ratio, and BAU. The common parameter space for the Yukawa couplings used for further analysis is provided in Table \ref{W2T5}.
 \begin{figure}[H]
		\centering
        \begin{subfigure}[b]{0.45\textwidth}
			\centering
			\includegraphics[width=9cm, height=6cm]{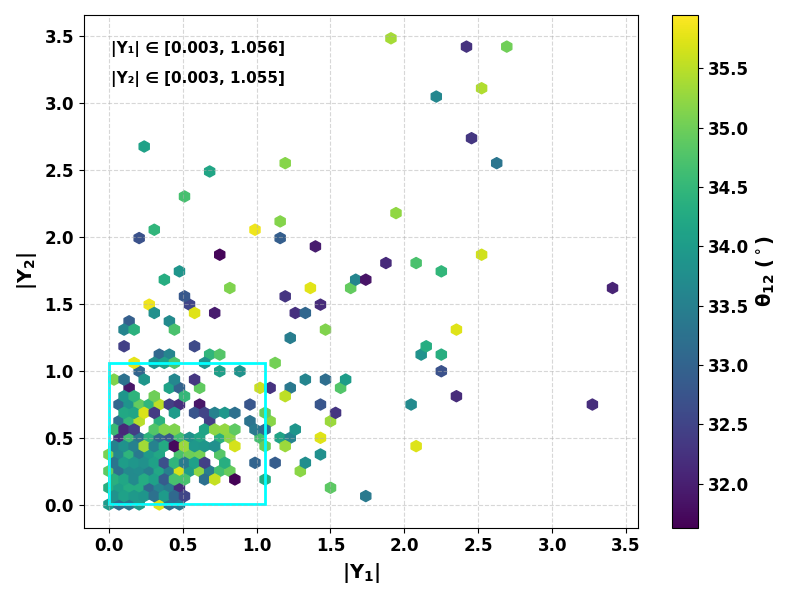}
			\caption{For NH}
			\label{p2f5}
		\end{subfigure}
		\hfill
		\begin{subfigure}[b]{0.45\textwidth}
			\centering
			\includegraphics[width=9cm, height=6cm]{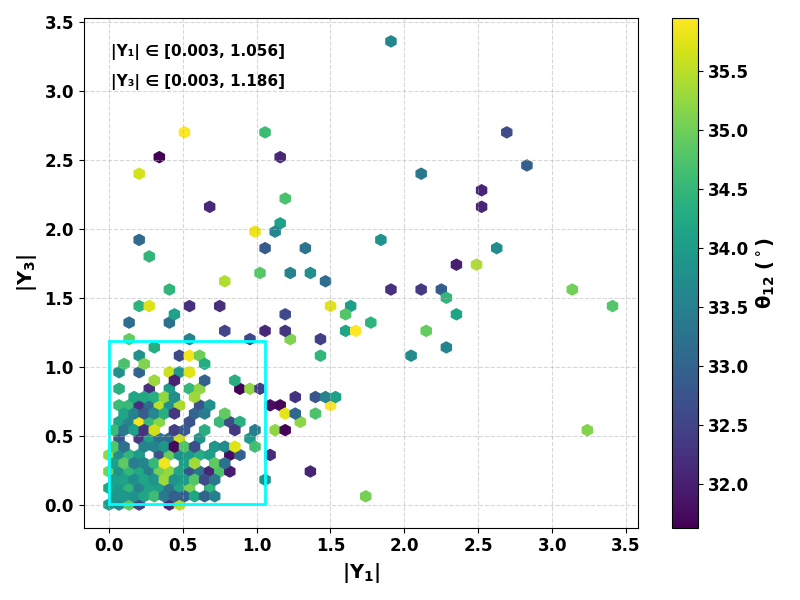}
			\caption{For NH}
			\label{p2f6}
		\end{subfigure}
	\begin{subfigure}[b]{0.45\textwidth}
			\centering
			\includegraphics[width=9cm, height=6cm]{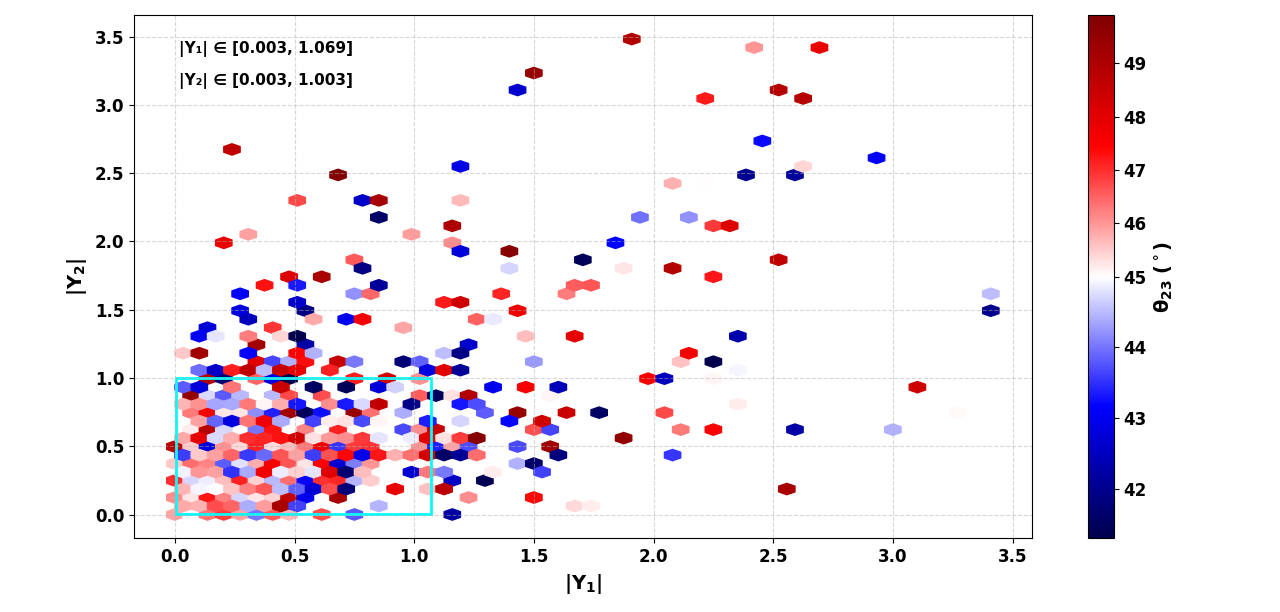}
			\caption{For IH}
			\label{p2f7}
		\end{subfigure}
		\hfill
		\begin{subfigure}[b]{0.45\textwidth}
			\centering
			\includegraphics[width=9cm, height=6cm]{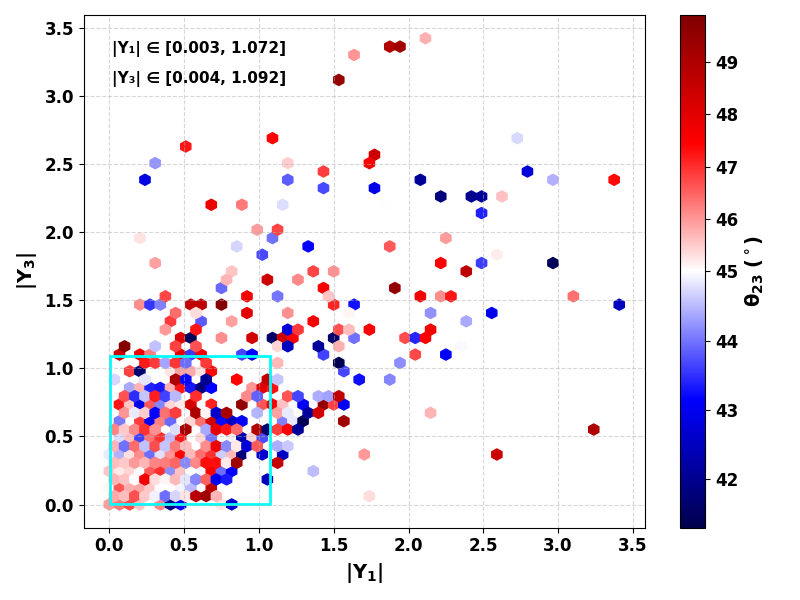}
			\caption{For IH}
			\label{p2f8}
		\end{subfigure}
		
		\vspace{0.5cm}
		
		\begin{subfigure}[b]{0.45\textwidth}
			\centering
			\includegraphics[width=9cm, height=6cm]{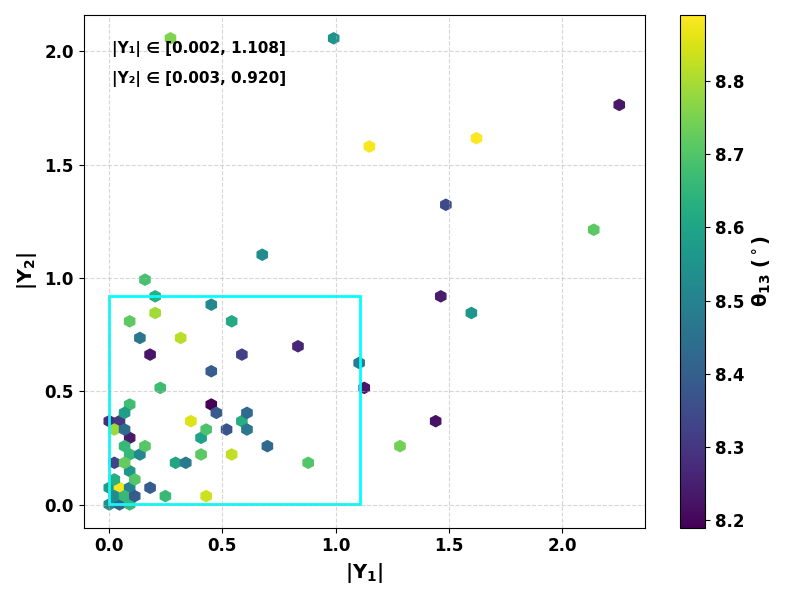}
			\caption{For NH}
			\label{p2f9}
		\end{subfigure}
		\hfill
		\begin{subfigure}[b]{0.45\textwidth}
			\centering
			\includegraphics[width=9cm, height=6cm]{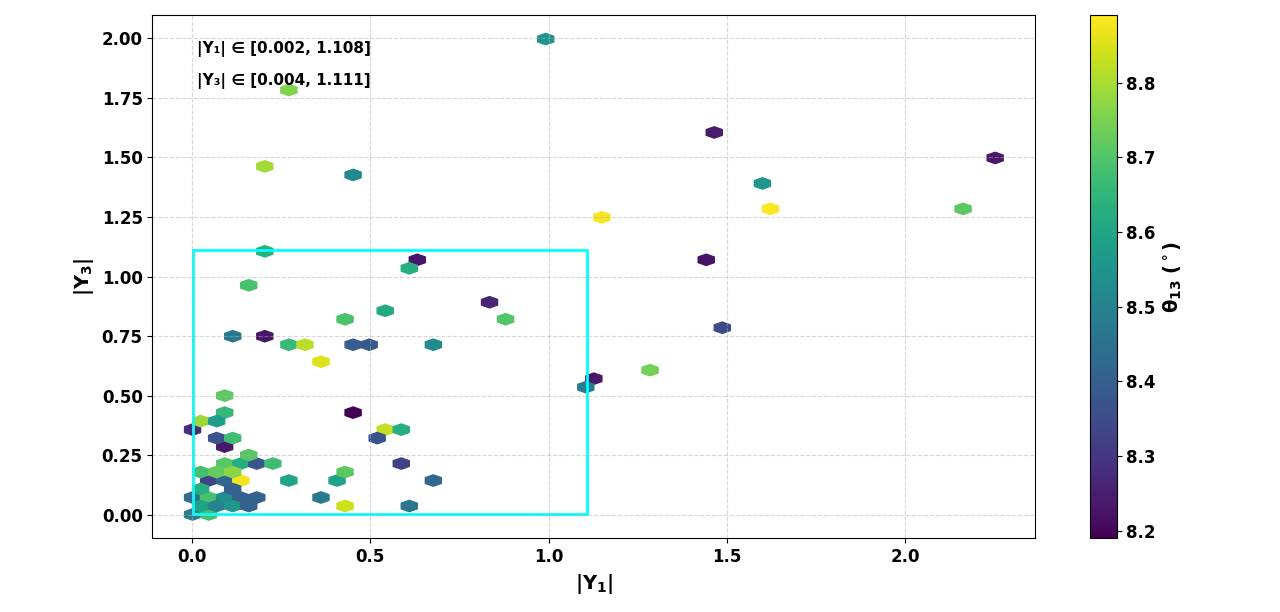}
			\caption{For NH}
			\label{p2f10}
		\end{subfigure}
		
		\caption{Density plots illustrating the parameter space of Yukawa couplings that satisfy the $3\sigma$ range of mixing angles for NH.}
		\label{P2F1}
	\end{figure}
		\begin{figure}[H]
		\centering
        \begin{subfigure}[b]{0.45\textwidth}
			\centering
			\includegraphics[width=9cm, height=6cm]{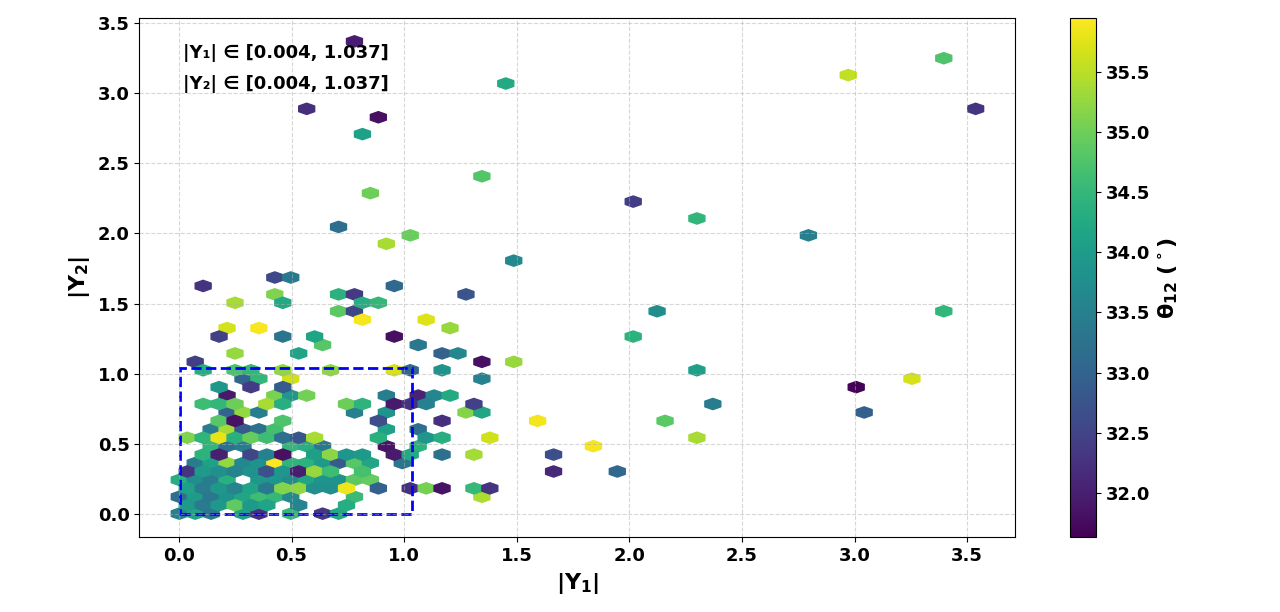}
			\caption{For NH}
			\label{p2f11}
		\end{subfigure}
		\hfill
		\begin{subfigure}[b]{0.45\textwidth}
			\centering
			\includegraphics[width=9cm, height=6cm]{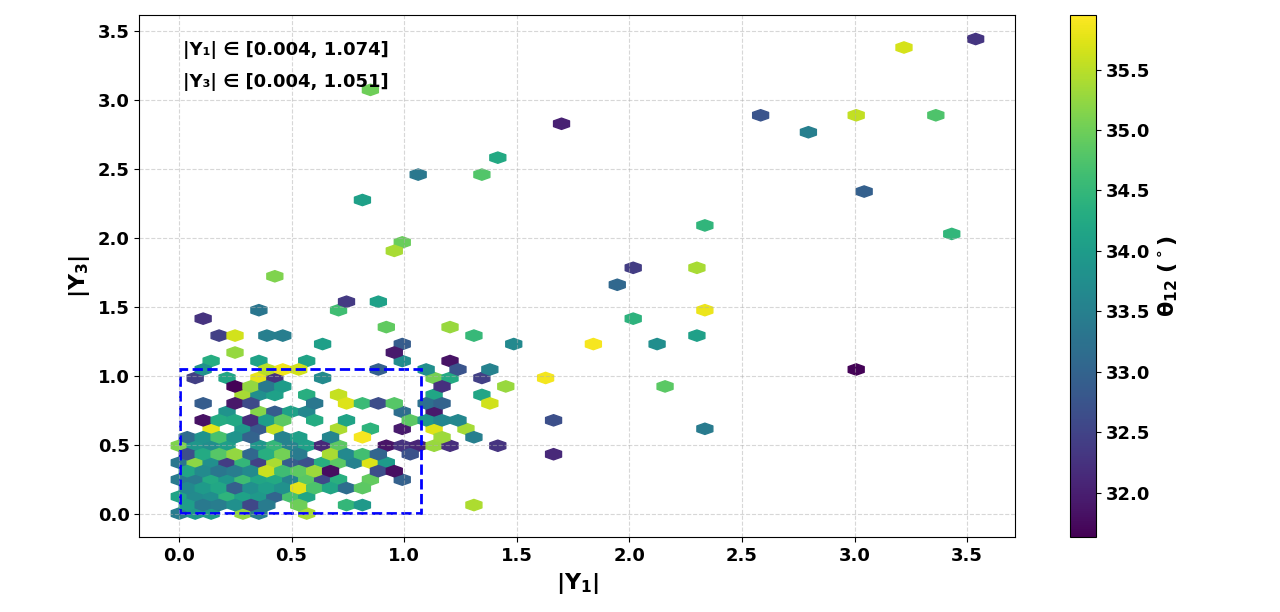}
			\caption{For NH}
			\label{p2f12}
		\end{subfigure}
        
		\begin{subfigure}[b]{0.45\textwidth}
			\centering
			\includegraphics[width=9cm, height=6cm]{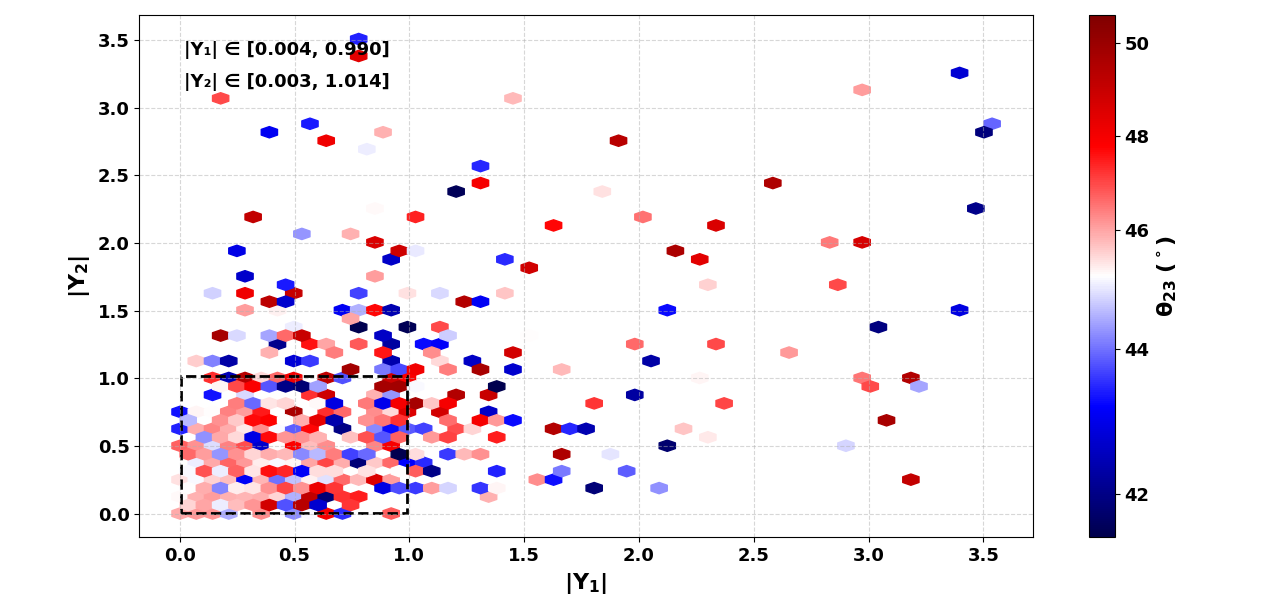}
			\caption{For IH}
			\label{p2f13}
		\end{subfigure}
		\hfill
		\begin{subfigure}[b]{0.45\textwidth}
			\centering
			\includegraphics[width=9cm, height=6cm]{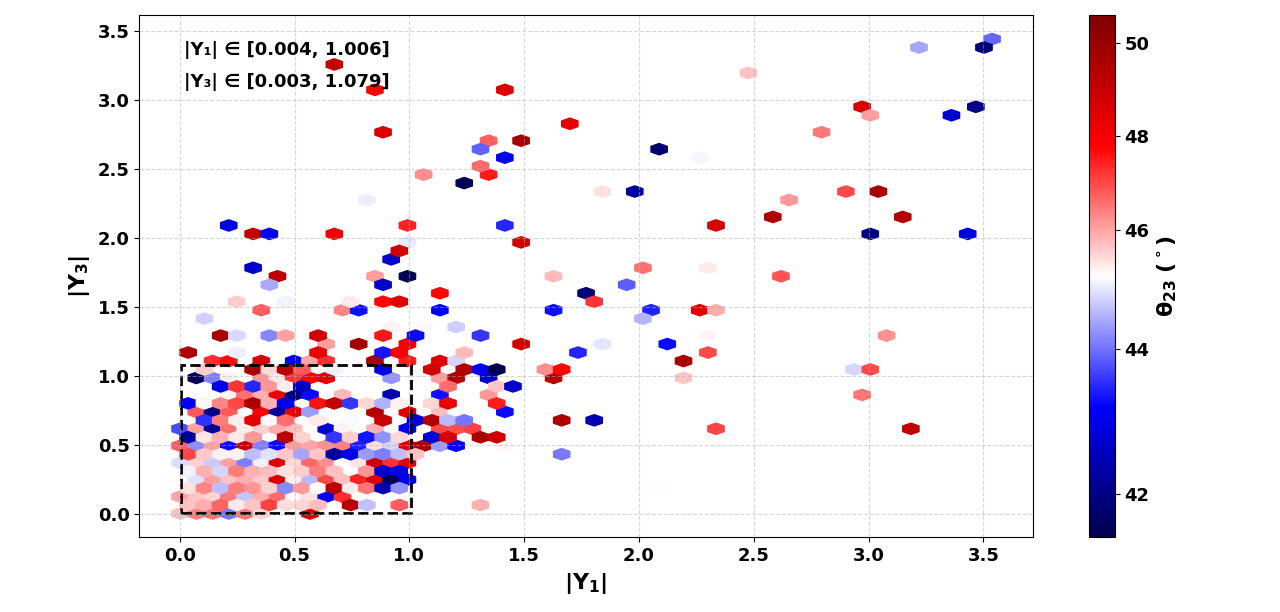}
			\caption{For IH}
			\label{p2f14}
		\end{subfigure}
		
		\vspace{0.5cm}
		
		\begin{subfigure}[b]{0.45\textwidth}
			\centering
			\includegraphics[width=9cm, height=6cm]{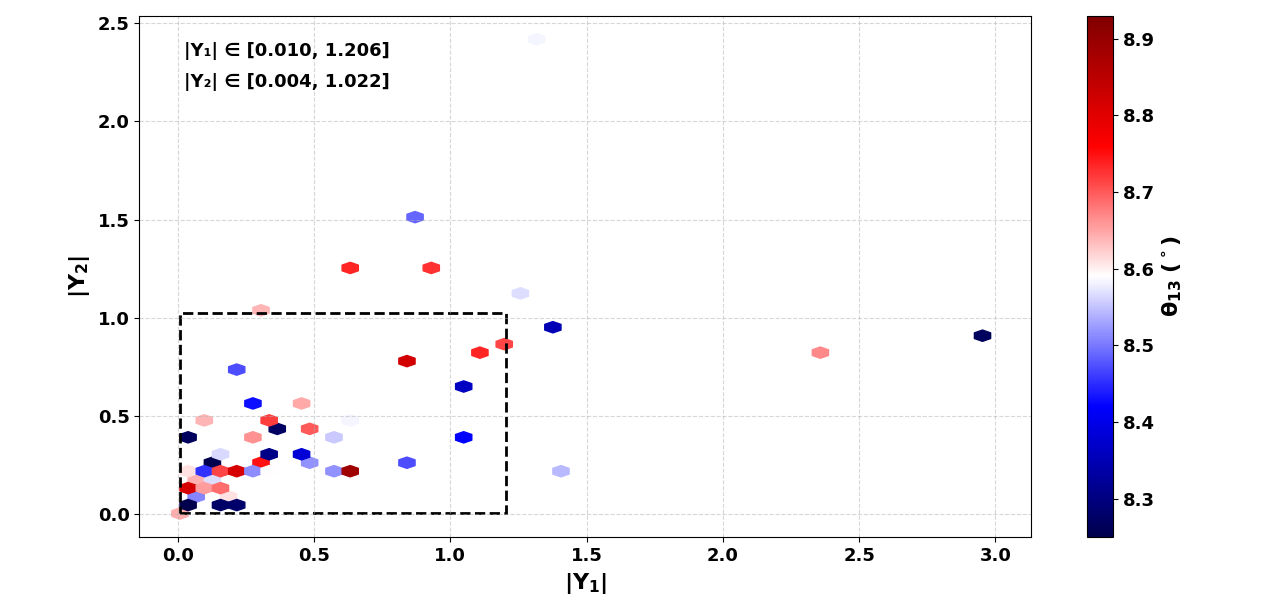}
			\caption{For NH}
			\label{p2f15}
		\end{subfigure}
		\hfill
		\begin{subfigure}[b]{0.45\textwidth}
			\centering
			\includegraphics[width=9cm, height=6cm]{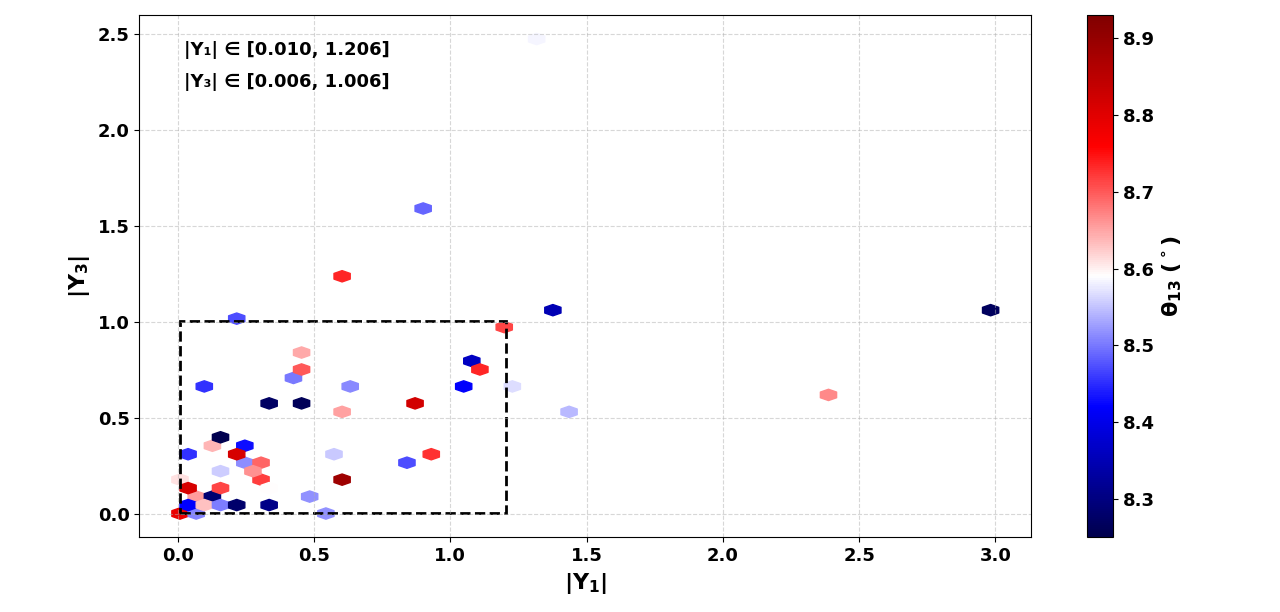}
			\caption{For NH}
			\label{p2f16}
		\end{subfigure}
		
		\caption{Density plots illustrating the parameter space of Yukawa couplings that satisfy the $3\sigma$ range of mixing angles for IH.}
		\label{P2F2}
	\end{figure}
    \begin{table}
		\centering
		\begin{tabular}{|c|c|c|}
			\hline
			Yukawa  coupling & for NH & for IH \\ \hline
			$Y_{1}$ & $0.003-1.056$  & $0.01-0.99$ \\ \hline
			$Y_{2}$ & $0.003-0.92$ &$0.004-1.014$ \\ \hline
			$Y_{3}$ &$0.003-1.092$ & $0.006-1.006$\\ \hline
		\end{tabular}
		\caption{Yukawa coupling values for NH and IH}
		\label{W2T5}
	\end{table}	
\subsection{Parameter space of modulus $\tau$}
\begin{figure}[h]
 	\centering
 	\begin{subfigure}[b]{0.45\textwidth}
 		\centering
 		\includegraphics[width=9cm, height=6cm]{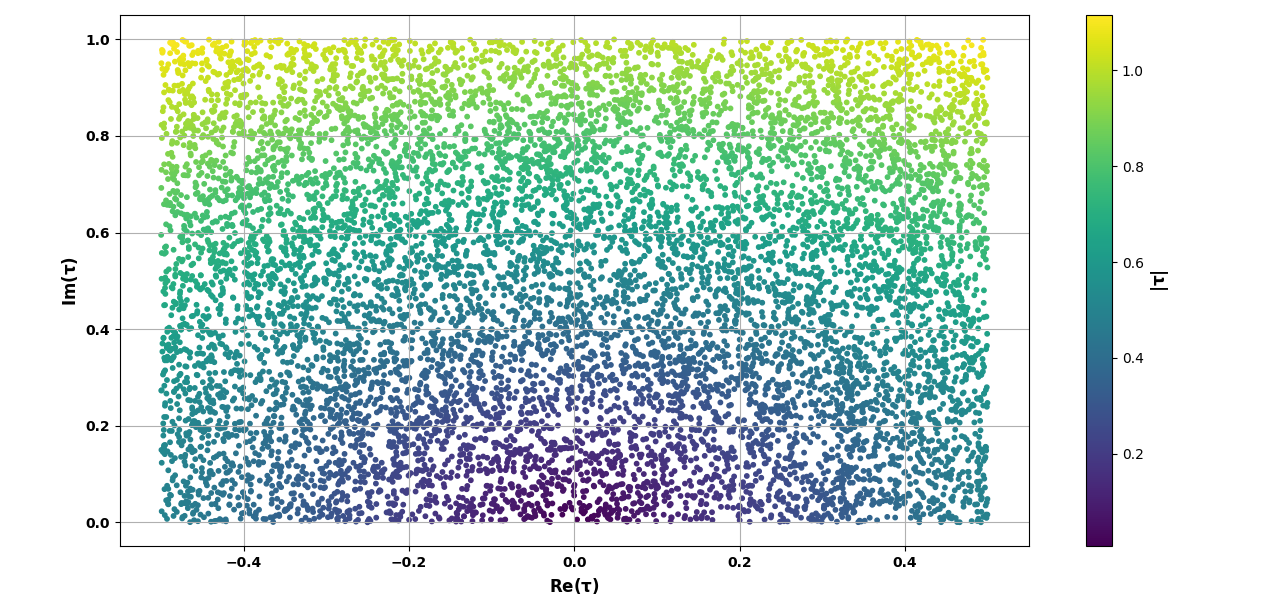}
 		\caption{For NH}
 		\label{p2f17}
 	\end{subfigure}
 	\hfill
 	\begin{subfigure}[b]{0.45\textwidth}
 		\centering
 		\includegraphics[width=9cm, height=6cm]{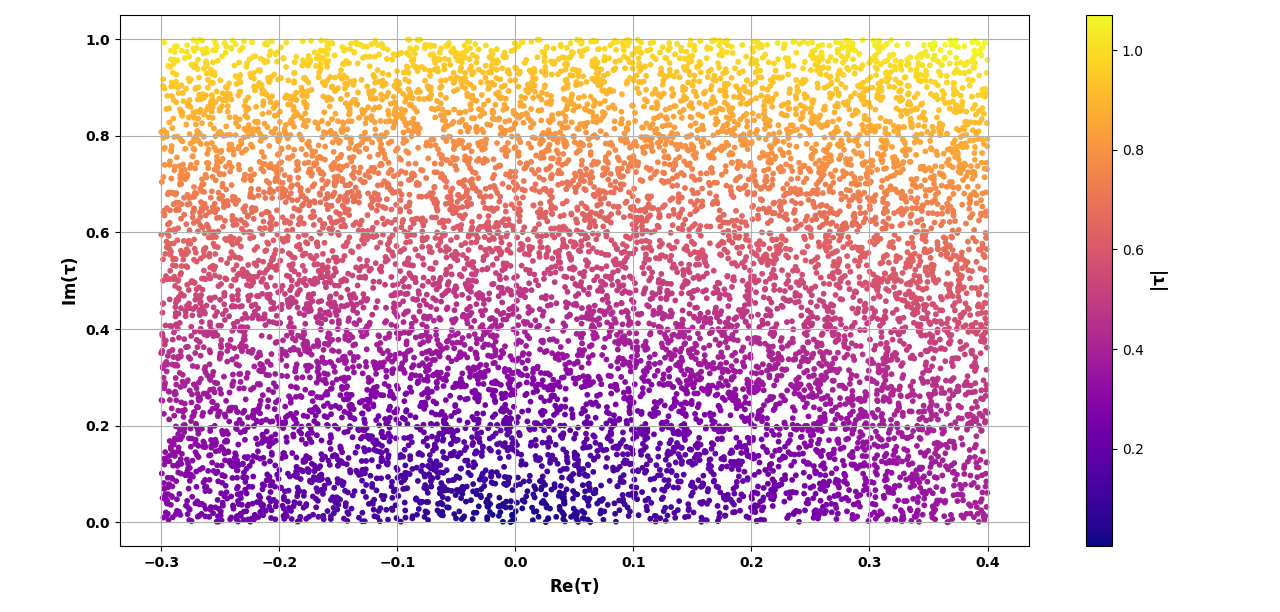}
 		\caption{For IH}
 		\label{p2f18}
 	\end{subfigure}
 	\caption{Figure show the parameter space of Re($\tau$) and Im($\tau$)}
 	\label{P2F3}
 \end{figure}

Within the model, we have a total of four modular forms, among which one is a singlet under $A_{4}$ and the other three form a triplet under $A_{4}$. The singlet modular form is a constant and is taken to be $1$ in our model. The $q$-expansions of the triplet modular forms are given in equation~\eqref{n4}. Since these modular forms correspond to the Yukawa couplings of the model, we compute the value of the modulus $\tau$ using their $q$-expansions after specifying a common range for the Yukawa couplings. To illustrate the parameter space of the real and imaginary parts of $\tau$, we have plotted the real part of $\tau$ against its imaginary part in Figure~\ref{P2F3}, for both NH and IH. We find that the real part of $\tau$ ranges from $-0.5$ to $0.5$, while the imaginary part of $\tau$ spans the full parameter space from $0$ to $1$ for NH. In case of IH, the real part of $\tau$ ranges from $-0.3$ to $0.4$ and the imaginary part of $\tau$ ranges from $0$ to $1$
\subsection{Calculation of effective Majorana mass}
\begin{figure}[h]
		\centering
		\begin{subfigure}[b]{0.45\textwidth}
			\centering
			\includegraphics[width=9cm, height=6cm]{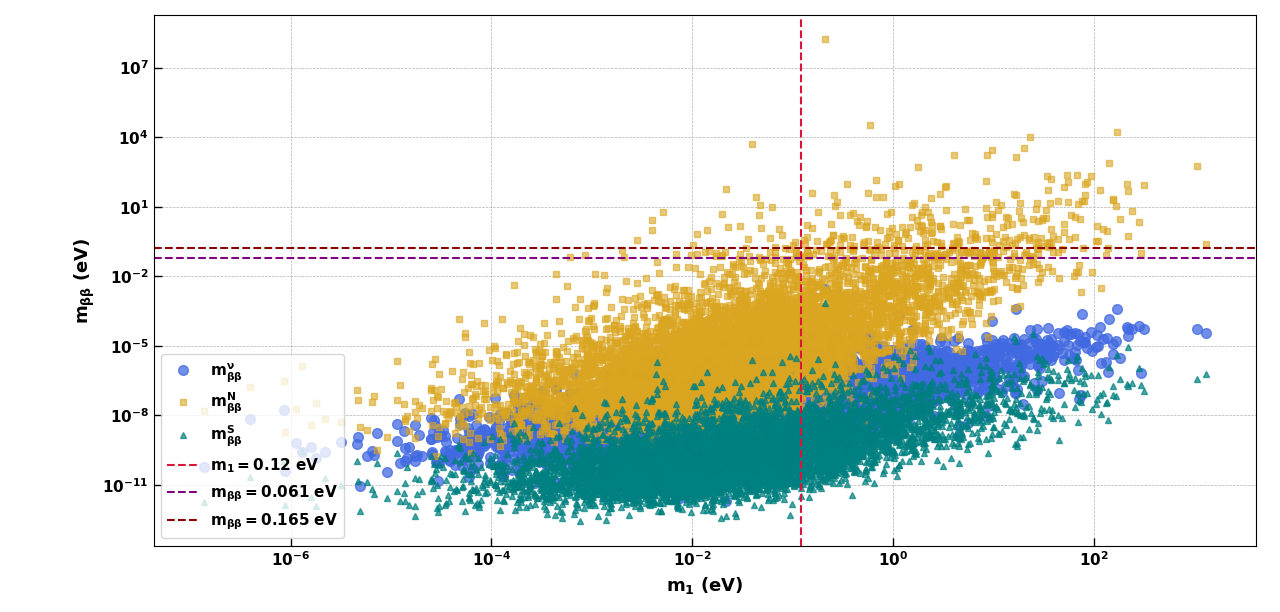}
			\caption{For NH}
			\label{p2f19}
		\end{subfigure}
		\hfill
		\begin{subfigure}[b]{0.45\textwidth}
			\centering
			\includegraphics[width=9cm, height=6cm]{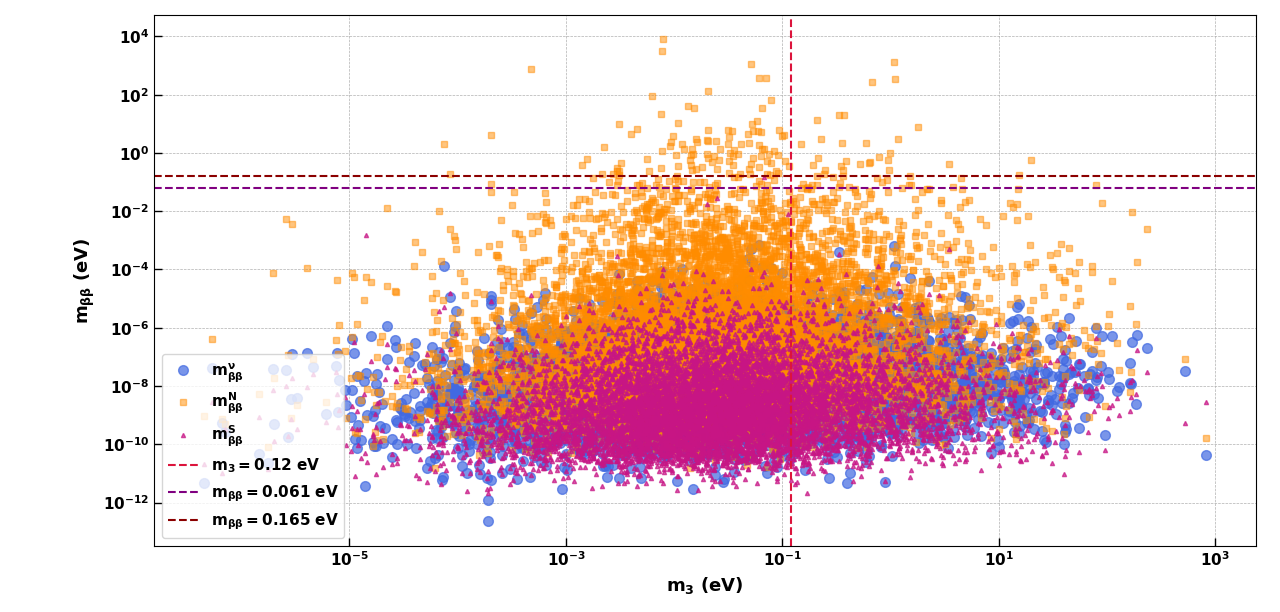}
			\caption{For IH}
			\label{p2f20}
		\end{subfigure}
		\caption{Figure shows the variation of effective Majorana mass with the lightest neutrino mass}
		\label{P2F4}
	\end{figure}

To see the contribution of the model to the $0\nu\beta\beta$ decay, we have calculated the effective mass for the $\lambda$ contribution. The formula that has been used in the calculation of effective mass is given in the equation \eqref{W2Q23}. The scattered plot in Figure \ref{P2F4} shows the variation of the effective mass with the lightest neutrino mass due to the exchange of light neutrino, heavy RH and sterile neutrino. The model contributes significantly to the calculation of effective mass as most of the calculated values fall below the experimental results.
\subsection{Calculation of Branching Ratio}
\begin{figure}[H]
		\centering
		\begin{subfigure}[b]{0.45\textwidth}
			\centering
			\includegraphics[width=9cm, height=6cm]{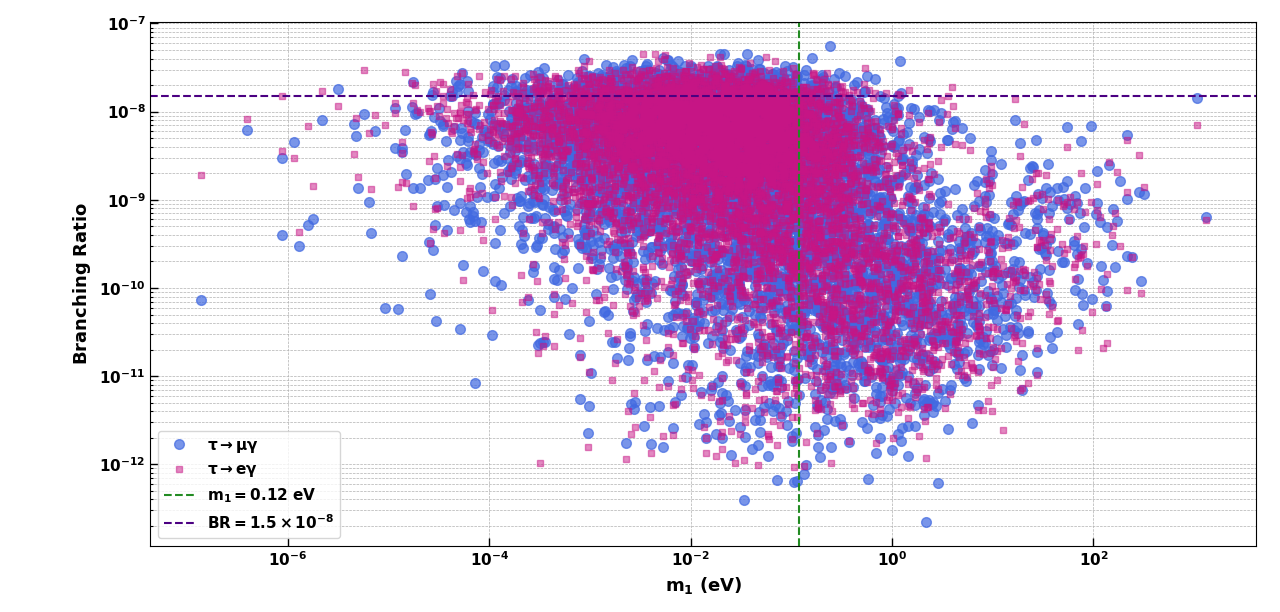}
			\caption{For NH}
			\label{p2f21}
		\end{subfigure}
		\hfill
		\begin{subfigure}[b]{0.45\textwidth}
			\centering
			\includegraphics[width=9cm, height=6cm]{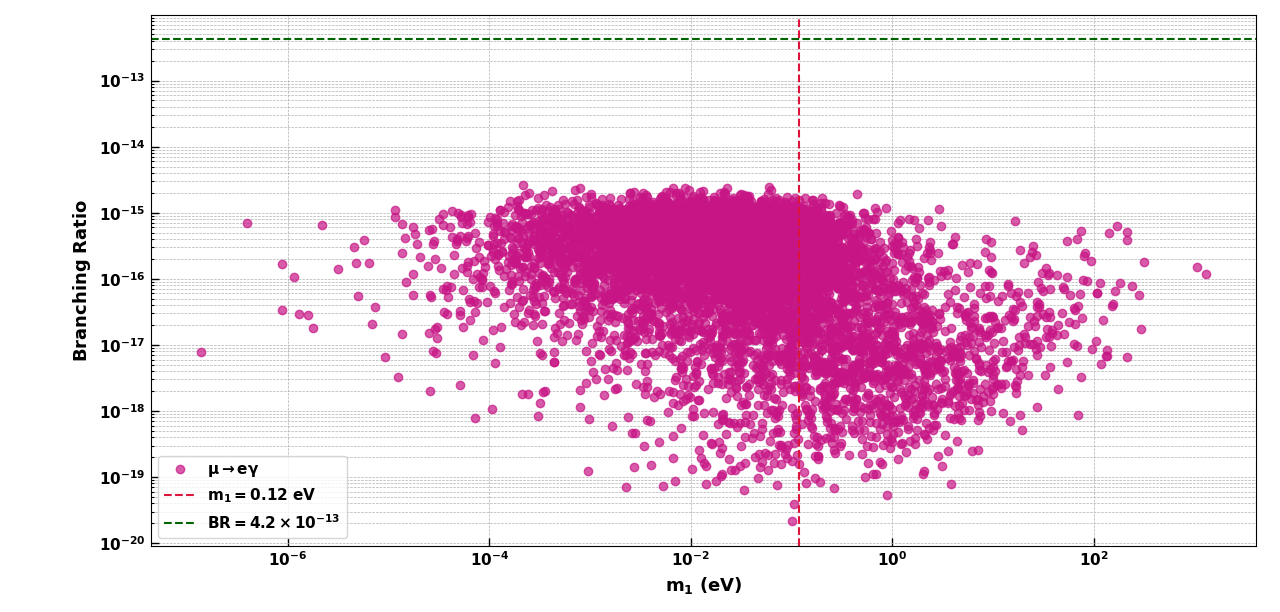}
			\caption{For NH}
			\label{p2f22}
		\end{subfigure}
		
		\vspace{0.5cm}
		
		\begin{subfigure}[b]{0.45\textwidth}
			\centering
			\includegraphics[width=9cm, height=6cm]{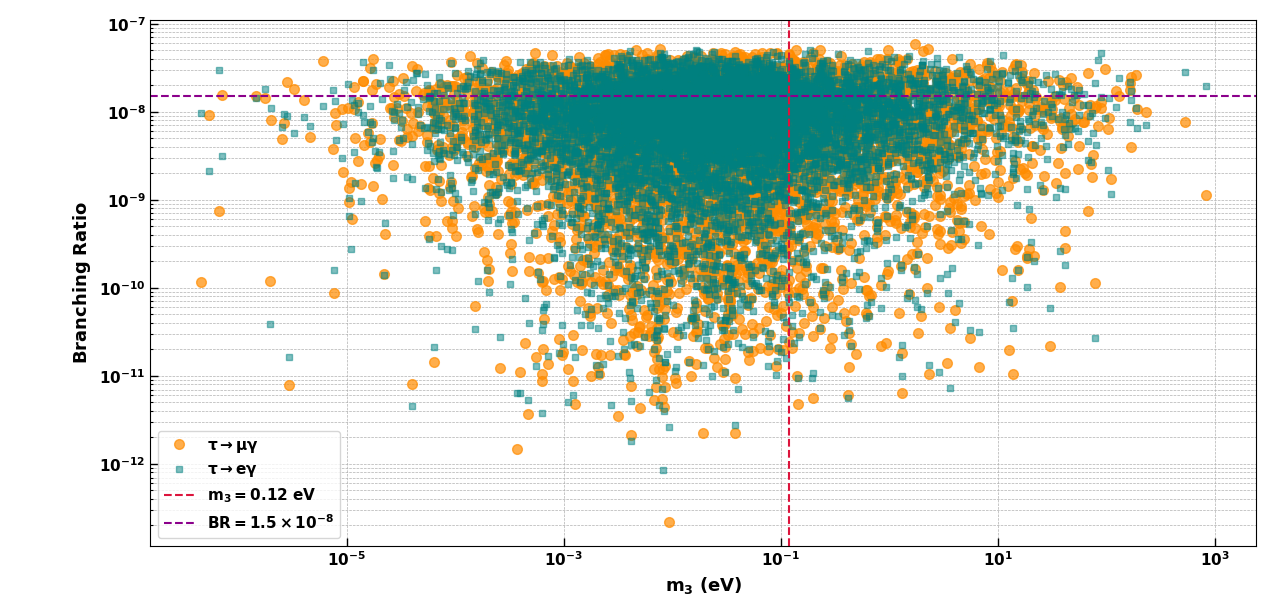}
			\caption{For IH}
			\label{p2f23}
		\end{subfigure}
		\hfill
		\begin{subfigure}[b]{0.45\textwidth}
			\centering
			\includegraphics[width=9cm, height=6cm]{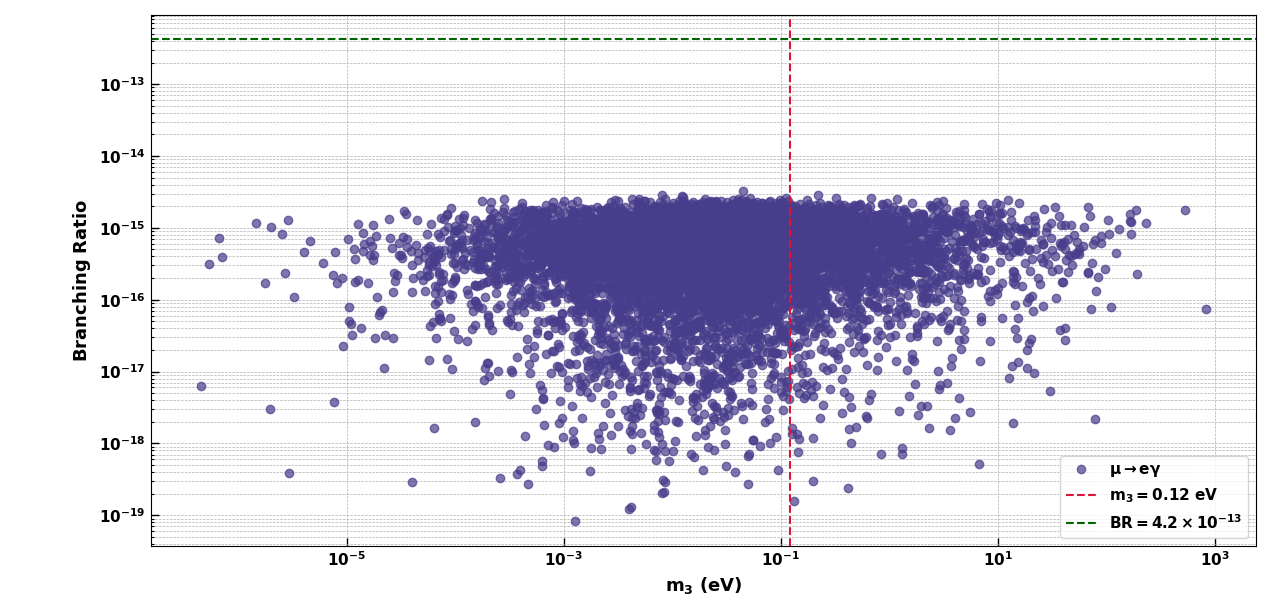}
			\caption{For IH}
			\label{p2f24}
		\end{subfigure}
		
		\caption{Figure shows the parameter space lightest neutrino mass that satisfies the bound on Branching ratio}
		\label{P2F5}
	\end{figure}

The branching ratio has been calculated using Equation~\eqref{W2Q22} for three lepton number violating processes, namely $\mu \rightarrow e + \gamma$, $\tau \rightarrow \mu + \gamma$, and $\tau \rightarrow e + \gamma$. The plots shown in Figure~\ref{P2F5} illustrate the variation of the branching ratio with respect to the lightest neutrino mass. In these figures, the horizontal lines represent the experimental bounds on the LFV decays, while the vertical lines indicate the upper bound on $\sum m_{\nu}$. \\
We observe that the maximum calculated branching ratios for the decays $\tau \rightarrow \mu + \gamma$ and $\tau \rightarrow e + \gamma$ lie below the current experimental limits, with most of the predicted values falling in the range of $10^{-8}$ to $10^{-11}$ for both NH and IH. However, in the case of the decay $\mu \rightarrow e + \gamma$, all the calculated values satisfy the experimental constraint, predicting branching ratios in the range of $10^{-15}$ to $10^{-18}$.
\subsection{Calculation of BAU}
\begin{figure}[h]
		\centering
		\begin{subfigure}[b]{0.45\textwidth}
			\centering
			\includegraphics[width=9cm, height=6cm]{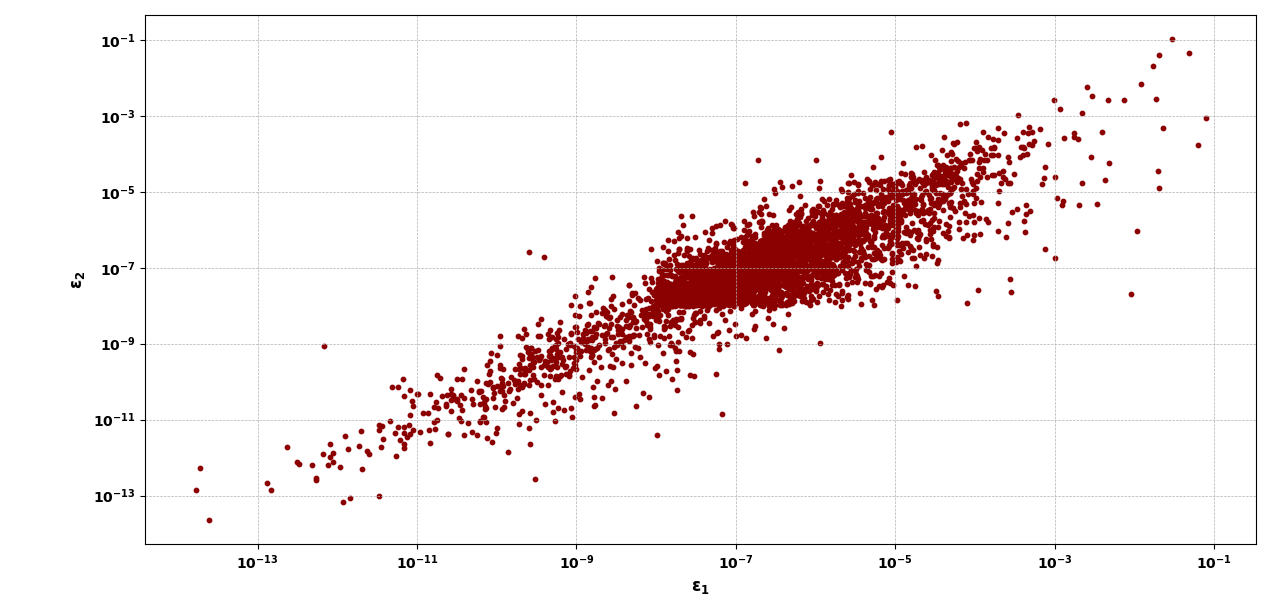}
			\caption{For NH}
			\label{p2f25}
		\end{subfigure}
		\hfill
		\begin{subfigure}[b]{0.45\textwidth}
			\centering
			\includegraphics[width=9cm, height=6cm]{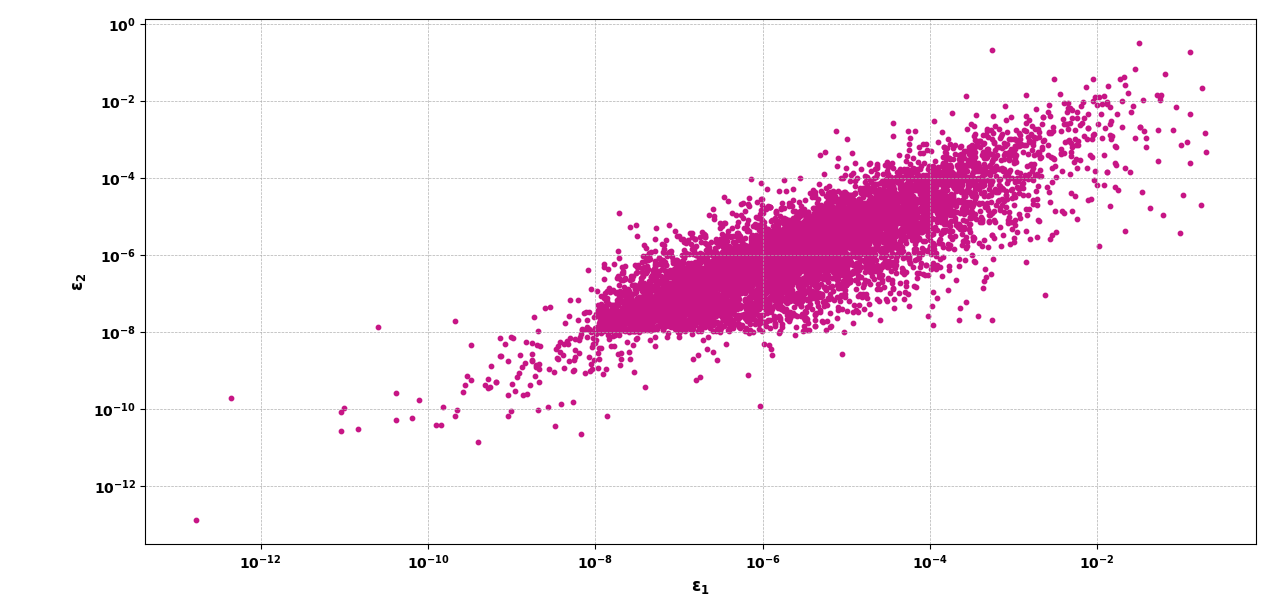}
			\caption{For IH}
			\label{p2f26}
		\end{subfigure}
		\caption{Figure shows the parameter space of CP asymmetry}
		\label{P2F4b}
	\end{figure}

\begin{figure}[H]
		\centering
		\begin{subfigure}[b]{0.45\textwidth}
			\centering
			\includegraphics[width=9cm, height=6cm]{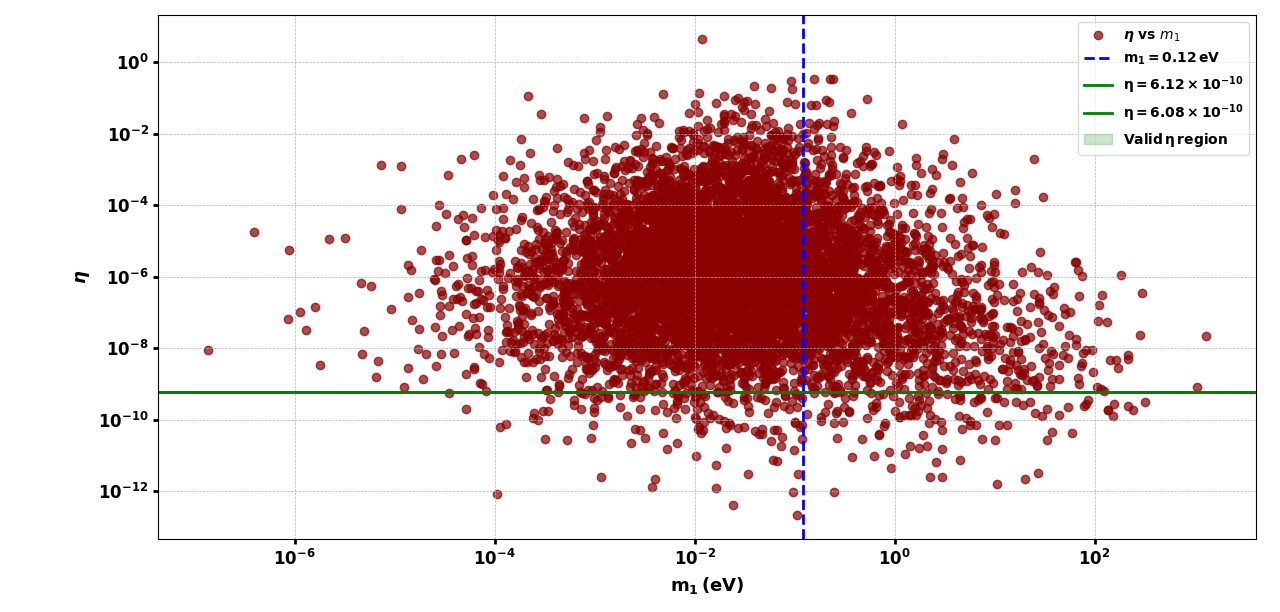}
			\caption{For NH}
			\label{p2f27}
		\end{subfigure}
		\hfill
		\begin{subfigure}[b]{0.45\textwidth}
			\centering
			\includegraphics[width=9cm, height=6cm]{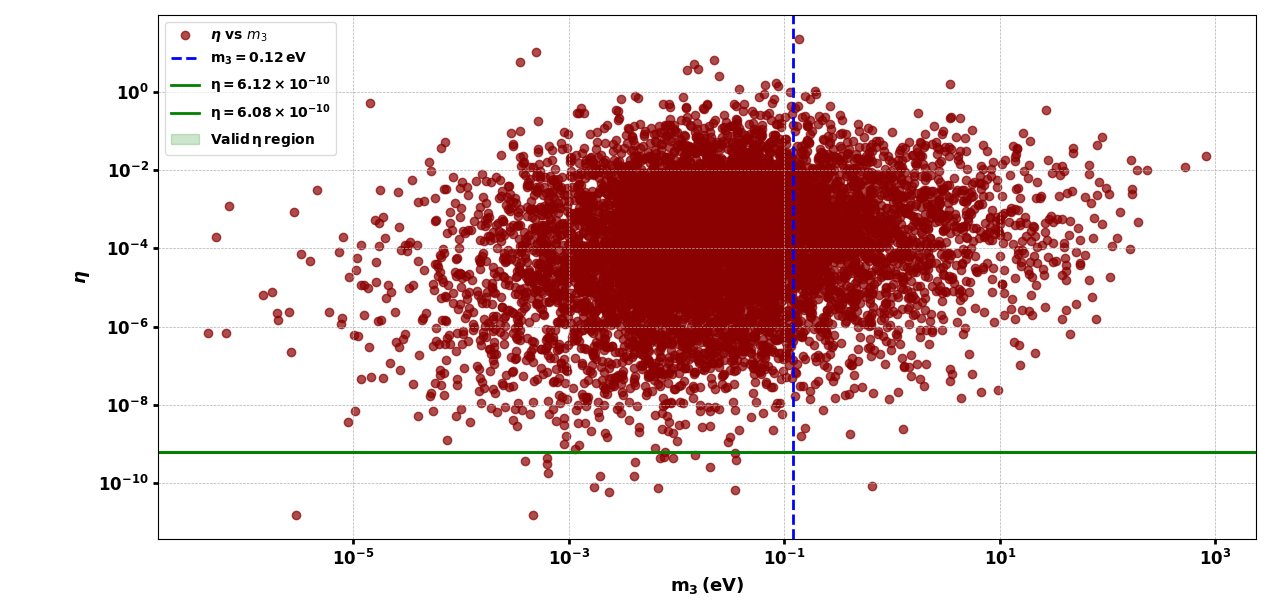}
			\caption{For IH}
			\label{p2f28}
		\end{subfigure}
		
		\vspace{0.5cm}
		
		\begin{subfigure}[b]{0.45\textwidth}
			\centering
			\includegraphics[width=9cm, height=6cm]{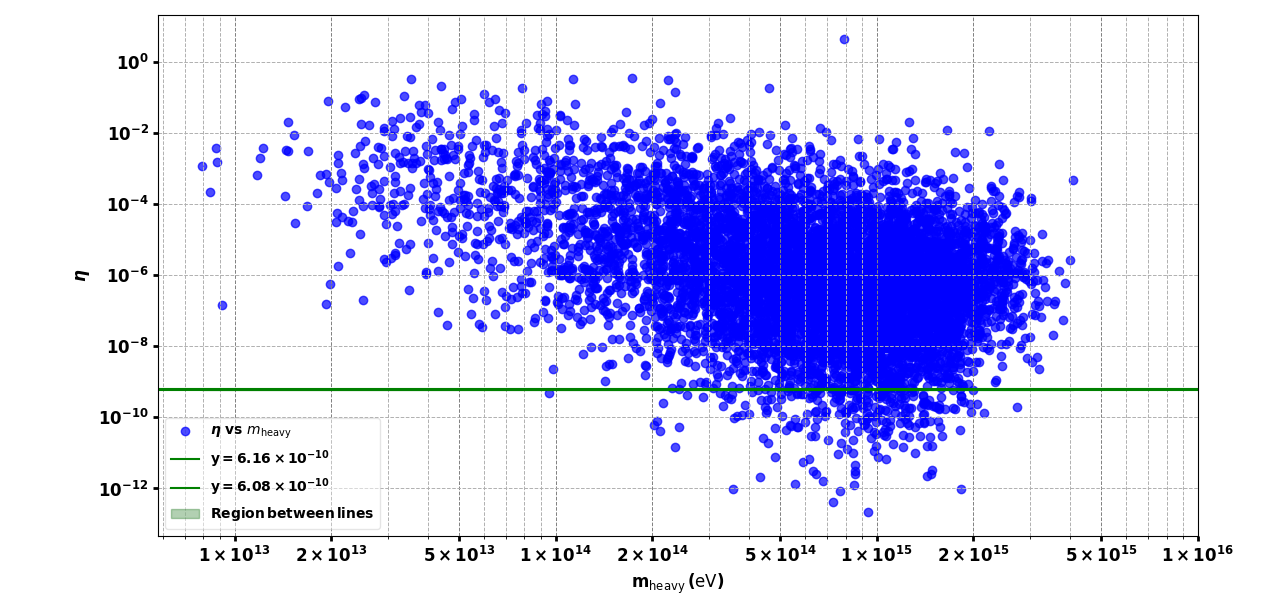}
			\caption{For NH}
			\label{p2f29}
		\end{subfigure}
		\hfill
		\begin{subfigure}[b]{0.45\textwidth}
			\centering
			\includegraphics[width=9cm, height=6cm]{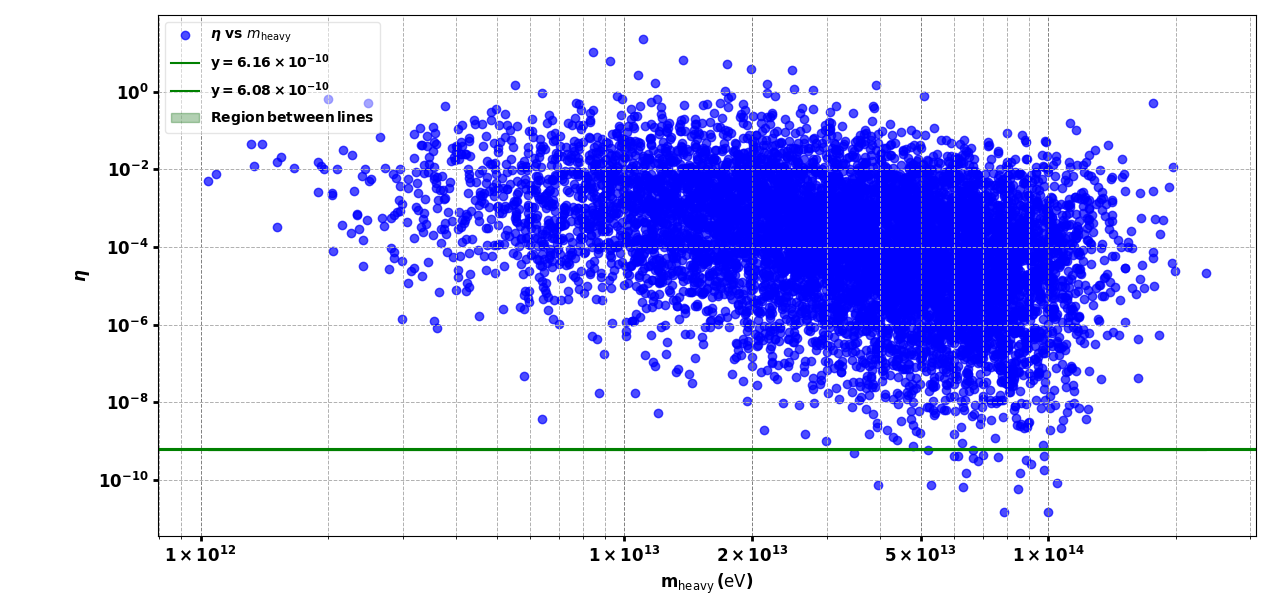}
			\caption{For IH}
			\label{p2f30}
		\end{subfigure}
		
		\caption{The top two figures illustrate the parameter space of the lightest neutrino mass that satisfies the BAU bound, while the bottom two figures depict the parameter space of the heavy neutrino mass that meets the BAU constraint.}
		\label{P2F6}
	\end{figure}

To study the model's capability to explain the baryon asymmetry of the universe, we attempt to generate the observed baryon asymmetry within the framework of our model. The model contains a total of six heavy neutrinos, whose mass eigenvalues are obtained by diagonalizing the $6 \times 6$ matrix given in Eq.~\eqref{W2Q25}. We find that the heavy mass eigenvalues are nearly degenerate, with mass splittings comparable to their decay widths. These conditions satisfy the condition for resonant leptogenesis, enabling the calculation of baryon asymmetry via this mechanism. Among the six heavy neutrinos, the pair $(\Psi_{5}, \Psi_{6})$ forms the lightest nearly degenerate pair. For this reason, we calculate the CP asymmetry term using the equation provided below.
\begin{gather} \label{W2Q33}
	\begin{aligned}
	\epsilon_{5} & =\frac{1}{8\pi (hh^{\dagger})_{55}}\text{Im}\Big[(hh^{\dagger})^{2}_{51}f_{51}+(hh^{\dagger})^{2}_{52}f_{52}+(hh^{\dagger})^{2}_{53}f_{53}+(hh^{\dagger})^{2}_{54}f_{54}+(hh^{\dagger})^{2}_{56}f_{56}\Big] \\
		\epsilon_{6} &= \frac{1}{8\pi (hh^{\dagger})_{66}}\text{Im}\Big[(hh^{\dagger})^{2}_{61}f_{61}+(hh^{\dagger})^{2}_{62}f_{62}+(hh^{\dagger})^{2}_{63}f_{63}+(hh^{\dagger})^{2}_{64}f_{64}+(hh^{\dagger})^{2}_{65}f_{65}\Big].
	\end{aligned}
\end{gather}
After computing the CP asymmetry, we employ Eq.~\eqref{W2Q30} to evaluate the washout factor ($K_{i}$), and the calculated values of $K_{i}$ lie in the range $10 \leq K_{i} \leq 10^{6}$. Subsequently, we determine the dilution factor $d$ using equation~\eqref{W2Q23}. Finally, the BAU is calculated using equation ~\eqref{W2Q31}.
Figure~\ref{P2F4b} shows the parameter space of the calculated values of the CP asymmetry term. We find that most of the values lie in the range of $10^{-7}$ to $10^{-3}$ for NH, while for IH, the values lie in the range of $10^{-8}$ to $10^{-2}$. We plot the calculated BAU values against the lightest neutrino mass for both NH and IH, as shown in Figures~\ref{p2f27} and~\ref{p2f28}, respectively.
In the case of NH, the asymmetry is observed when the lightest neutrino mass varies from $ 0.12$~eV down to approximately $10^{-4}$~eV. However, for IH, only a few data points lie within the allowed BAU region.
Figures~\ref{p2f29} and~\ref{p2f30} show the plots of heavy neutrino mass versus the BAU. For NH, the BAU bound is satisfied when the heavy neutrino mass lies between $200$~TeV and $2000$~TeV. In contrast, for IH, the BAU bound is satisfied when the heavy neutrino mass ranges from $50$~TeV to $100$~TeV.
\section{Conclusion}\label{W2S7} In this study, we have constructed a non-supersymmetric model using modular symmetry and explored various BSM phenomena. We have calculated the effective Majorana mass from the $0\nu\beta\beta$ process due to the contribution of the $W_{L}-W_{R}$ current. Additionally, we have computed the branching ratios for LFV processes. To further assess the model's predictivity, we have studied the BAU within the framework of resonant leptogenesis. The main findings of this work can be summarized as follows:
\begin{itemize}
\item We have developed a non-supersymmetric model based on modular symmetry. While modular symmetry has been predominantly used in supersymmetric neutrino mass models, where the Yukawa couplings are holomorphic functions of the modulus $\tau$, our study considers a non-supersymmetric framework. In this case, the Yukawa couplings can include both holomorphic and non-holomorphic components. We have implemented the $\Gamma_{3}$ modular group to realize a non-supersymmetric left-right asymmetric model. Instead of the conventional type-I or type-II seesaw mechanisms, we have employed an extended inverse seesaw mechanism to generate the light neutrino masses. We have constructed the Dirac, Majorana, $N-S$, and $S-S$ mixing mass matrices in terms of three Yukawa couplings ($Y_{1}$, $Y{2}$, $Y_{3}$). After determining the Yukawa coupling values from the model, we performed density plots to identify the parameter space of Yukawa coupling for which the mixing angles fall within experimental constraints. The common range for these Yukawa couplings is found to be in the order of $10^{-2}$ to $1$ for both NH and IH.
		
\item We have calculated the effective Majorana mass due to the $W_{L}-W_{R}$ current. The obtained values of the effective mass lie below the experimental bounds. We observe that the effective mass remains very small when the exchange particle is either active neutrinos or sterile neutrinos for both NH and IH cases. However, when the exchange particle is heavy RH neutrino, the effective mass ranges from approximately $10^{-1}$ eV to $10^{-6}$ eV. 
Thus, in the present model, the effective mass remains small when mediated by active or sterile neutrinos, whereas a significant contribution is observed when a RH neutrino acts as the exchange particle. Although some calculated values exceed the experimental bounds, the majority remain within acceptable limits, demonstrating the model's viability in predicting the effective mass.
		
\item We have computed the branching ratios for three LFV processes. For all three processes, the predicted branching ratios are found to be consistent with the experimental limits.
		
\item To further verify the model's predictivity, we have investigated the baryon asymmetry of the universe. In our model, the asymmetry is generated via the decay of a quasi-Dirac pair of heavy neutrinos. For NH, a successful BAU is achieved when the mass eigenvalues of the heavy neutrinos range from $200$ TeV to $2000$ TeV. In the IH case, we find a very small number of parameter points satisfying the Planck constraint on BAU when the heavy neutrino masses range from $50$ TeV to $100$ TeV.
		
\item While analyzing the above phenomena, we have also examined the parameter space for the light neutrino masses that satisfy experimental constraints. Specifically, we have identified a common range of light neutrino masses capable of explaining all the above mentioned BSM phenomena. The light neutrino mass varies between $ 10^{-5}$ eV and $0.1$ eV for both NH and IH.
\end{itemize}
	
Overall, our study demonstrates that a non-supersymmetric model based on modular symmetry can successfully explain various neutrino properties and BSM phenomena while remaining consistent with experimental constraints.
\revappendix
\section{\textbf{Modular Group and Polyharmonic $Maa\beta$ form}} \label{W2A1}
The modular group $SL(2,Z)=\Gamma$ is defined as a group of $2\times2$ matrices with positive or negative integer element and determinant equal to $1$ and it represents the symmetry of a torus \cite{deAnda:2023udh,deAdelhartToorop:2011re}. It is infinite group and generated by two generators of the group $S$ and $T$.\\
	\begin{equation}\label{Q1}
		\Gamma = \Biggl\{\begin{pmatrix}
			a & b\\
			c & d
		\end{pmatrix} | a,b,c,d \in \mathbb{Z}, ad-bc=1\Biggr\}~~ .
	\end{equation}
	The generator of the group satisfy the conditions:\\
	$S^{2} = 1$ \hspace{0.5cm} and  \hspace{0.5cm} $(ST)^{3} = 1$ \\
	and they can be represented by $2\times 2$ matrices 
	\begin{align*}
		S=\begin{pmatrix} 
			0 & 1\\
			-1 & 0 \\
		\end{pmatrix},
		&  ~~~~~~
		T= \begin{pmatrix}
			1 & 1 \\
			0 & 1 \\
		\end{pmatrix} ~~~.
	\end{align*}
	A two-dimensional space is obtained, when the torus is cut open and this two-dimensional space can be viewed as an Argand plane and modulus $\tau$ is the lattice vector of that Argand plane. The transformation of modulus $\tau$ \cite{Ferrara:1989qb} of the modular group on the upper half of the complex plane is given below 
	\begin{equation*}
		\gamma : \tau \rightarrow \gamma(\tau) = \frac{(a\tau + b)} {(c\tau + d)}
	\end{equation*} 
	the transformation of $\tau$ is same for both $\gamma$ and $-\gamma$ and we can define a group $\bar{\Gamma}= PSL(2,Z)$, which is a  projective special linear group. Also, the modular group has an infinite number of normal subgroups, which is the principal congruence subgroup of level N and can be defined as
	\begin{equation}\label{Q2}
		\Gamma(N)= \Biggl\{\begin{pmatrix}
			a & b\\
			c & d
		\end{pmatrix} \in SL(2,Z), \begin{pmatrix}
			a & b \\
			c & d
		\end{pmatrix}= \begin{pmatrix}
			1 & 0 \\
			0 & 1
		\end{pmatrix} (mod N)\Biggr\},
	\end{equation}  
	and for $N>2$, $\bar{\Gamma}(N)=\Gamma_{N}$. In model building purpose it is efficient to use finite group. Usually a modular group is an infinite group but we can obtain a finite modular group for $N>2$, if we consider the quotient group $\Gamma_{N} =PSL(2,Z)/\bar{\Gamma}(N)$ and those modular group are isomorphic to non-abelian discrete groups. \\
	The modular invariance required the Yukawa couplings to be a certain modular function $Y(\tau)$ and should fellow the following transformation property.
	
	\begin{equation}\label{n2}
		Y(\gamma\tau)=(c\tau+d)^{k}Y(\tau)
	\end{equation} 
	One can realise the non-supersymmetric framework by using the framework of automorphic form and the assumption of holomorphicity is replaced by the Laplacian condition and in such case the Yukawa coupling can have both holomorphic and non-holomorphic parts \cite{Qu:2024rns,Ding:2020zxw,Ding:2024inn}. In the present work, we are concerned with the polyharmonic $Maa\beta$ forms of weight k and the Yukawa coupling needs to follow another transformation property, which is given below
	\begin{equation}\label{n3}
		\Delta_{k}Y(\tau)=0
	\end{equation}
	where $\tau=x+iy$ and $\Delta_{k}$ is the hyperbolic Laplacian operator
	\begin{equation}
		\Delta_{k}=-y^{2}\big(\frac{\partial^{2}}{\partial x^{2}} + \frac{\partial^{2}}{\partial y^{2}}\big) + iky\big(\frac{\partial}{\partial x}+ i\frac{\partial}{\partial y}\big) = -4y^{2} \frac{\partial}{\partial \tau}\frac{\partial}{\partial \tau}+ 2iky \frac{\partial}{\partial \tau}
	\end{equation}
	The weight k of polyharmonic $Maa\beta$ forms can be positive, zero, or even negative. Based on the transformation property given in the equation \eqref{n2}, which implies that $Y(\tau + N)=Y(\tau)$ and considering the transformation property from equation \eqref{n3}, the Fourier expansion of a level N and weight k polyharmonic $Maa\beta$ form can be expressed as\cite{Qu:2024rns}
	\begin{equation}\label{n01}
		Y(\tau)= \sum_{n\in\frac{1}{N}\mathbb{Z} n\leqq 0} c^{+}(n)q^{n} + c^{-}(0)y^{1-k} + \sum_{n\in \frac{1}{N}\mathbb{Z}n\leq 0} c^{-}(n)\Gamma(1-k,-4\pi ny)q^{n}
	\end{equation}
	where $q=e^{i2\pi\tau}$
	In the present work, since we are working in the non-supersymmetric framework, we have focused on polyharmonic $Maa\beta$ forms of weight zero at level $3$. It can be arranged into a singlet and triplet under $A_{4}$ group. The q expansion of the weight zero Yukawa couplings at level three is provided below
	\begin{gather}\label{n4}
		\begin{aligned}
			Y^{(0)}_{3,1} &= y- \frac{3 e^{-4 \pi y}}{\pi q}- \frac{9 e^{-8\pi y}}{2\pi q^{2}}-\frac{-12\pi y}{\pi q^{3}}-\frac{21 e^{-16\pi y}}{4\pi q^{4}}-\frac{18 e^{-20\pi y}}{5\pi q^{5}}-\frac{3e^{-24\pi y}}{2\pi q^{6}}+\cdot\cdot\cdot\cdot\cdot \\& -\frac{9\log 3}{4\pi}-\frac{3q}{\pi}-\frac{9q^{2}}{2\pi}-\frac{q^{3}}{\pi}-\frac{21q^{4}}{4\pi}-\frac{18q^{5}}{5\pi}-\frac{3q^{6}}{2\pi} \\
			Y^{(0)}_{3,2} &= \frac{27q^{\frac{1}{3}}e^{\frac{\pi y}{3}}}{\pi}\big( \frac{e^{-3\pi y}}{4q} + \frac{e^{-7\pi y}}{5q^{2}} + \frac{5e^{-11\pi y}}{16q^{3}} + \frac{2e^{-15\pi y}}{11q^{4}} + \frac{2e^{-19\pi y}}{7q^{5}} + \frac{4e^{-23\pi y}}{17 q^{6}}+\cdot\cdot\cdot\cdot\cdot\cdot\cdot \big)\\& + \frac{9q^{\frac{1}{3}}}{2\pi}\big( 1+ \frac{7q}{4} + \frac{8q^{2}}{7}+ \frac{9q^{3}}{5}+\frac{14q^{4}}{13}+\frac{31q^{5}}{16}+\frac{20q^{6}}{19}+\cdot\cdot\cdot\cdot\big) \\
			Y^{(0)}_{3,3} &= \frac{9q^{\frac{2}{3}}e^{\frac{2\pi y}{3}}}{2\pi} \big(\frac{e^{-2\pi y}}{q}+\frac{7e^{-6\pi y}}{4q^{2}}+\frac{8e^{-10\pi y}}{7q^{3}}+ \frac{9e^{-14\pi y}}{5q^{4}}+\frac{14e^{-18\pi y}}{13q^{5}}+\frac{31e^{-22\pi y}}{16q^{6}}+\cdot\cdot\cdot\cdot\cdot\cdot \big) \\& +\frac{27q^{\frac{2}{3}}}{\pi}\big(\frac{1}{4}+\frac{q}{5}+\frac{5q^{2}}{16}+\frac{2q^{3}}{11}+\frac{2q^{4}}{7}+\frac{9q^{5}}{17}+\frac{21q^{6}}{20}+\cdot\cdot\cdot\cdot\cdot \big)
		\end{aligned}    
	\end{gather}
	\bibliographystyle{unsrt}
	\bibliography{cite}
	
\end{document}